\documentclass[final,5p,times,twocolumn]{elsarticle}

\usepackage{amsmath}
\usepackage{amssymb}
\usepackage{lipsum}  
\usepackage{placeins}
\usepackage{subcaption}
\usepackage{overpic}
\usepackage{url}
\usepackage[pdftex, pdftitle={Article}, pdfauthor={Author},colorlinks = true,allcolors = blue]{hyperref}

\usepackage{lineno}
\journal{NIMA}

\begin{document}

\affiliation[UCR]{Department of Physics and Astronomy, University of California, Riverside, CA 92521, USA}
\affiliation[JLAB]{Thomas Jefferson National Accelerator Facility, Newport News, VA 23606, USA}

\author[UCR,JLAB]{Miguel Arratia\corref{correspondingauthor}}
\cortext[correspondingauthor]{Corresponding author}

\author[UCR]{Ryan Milton}
\author[UCR,JLAB]{Sebouh Paul}
\author[UCR]{Barak Schmookler}
\author[UCR]{Weibin Zhang}

\title{A Few-Degree Calorimeter for the future Electron-Ion Collider}


\begin{abstract}
Measuring the region $0.1 < Q^{2} < 1.0$ GeV$^{2}$ is essential to support searches for gluon saturation at the future Electron-Ion Collider. Recent studies have revealed that covering this region at the highest beam energies is not feasible with current detector designs, resulting in the so-called $Q^{2}$ gap.  In this work, we present a design for the Few-Degree Calorimeter (FDC), which addresses this issue. The FDC uses SiPM-on-tile technology with tungsten absorber and covers the range of $-4.6 < \eta < -3.6$. It offers fine transverse and longitudinal granularity, along with excellent time resolution, enabling standalone electron tagging. Our design represents the first concrete solution to bridge the $Q^{2}$ gap at the EIC.

\end{abstract}

\maketitle
\newpage
\tableofcontents
\section{Introduction}
\label{sec:outline}

The technology currently proposed to cover the smallest electron-scattering angles with the central detector of the future Electron-Ion Collider (EIC) involves the use of lead-tungsten crystals~\cite{AbdulKhalek:2021gbh}. Although these crystals provide exceptional energy resolution, they face limitations when employed near the beampipe. The beampipe has a complex geometry due to the 25 mrad beam-crossing angle and incorporates flanges connecting the electron and hadron beampipes. Furthermore, the system loads necessitate a support structure near the beampipe with adequate clearance margins, which further limits their coverage.

The EIC Yellow Report~\cite{AbdulKhalek:2021gbh} established an acceptance requirement of $-4<\eta<4$. However, it is not feasible to meet the lower limit of this range using only a crystal calorimeter. The current design of the EIC detector, ePIC, places the crystal ECAL at $z = -174$ cm, with an inner radius of $\approx$8 cm~\cite{managerie,ELKEprivate}, resulting in a nominal acceptance that reaches $\eta\approx-3.6$. This limitation is mainly due to the assembly requirements of the ECAL, which necessitate a hole large enough to allow it to slide through the flange further downstream. 

At the highest EIC energy, $\eta=-3.6$ corresponds to $Q^{2}\approx$ 1 GeV$^{2}$. Consequently, the limited acceptance of the ECAL hinders the study of the transition to the perturbative QCD domain, for which measuring $Q^{2}<$ 1 GeV$^{2}$ is necessary.  This limitation affects various physics topics at the EIC, including flagship studies such as searches for gluon saturation.

The region of very low $Q^{2}$ will be covered by far-backward taggers~\cite{AbdulKhalek:2021gbh}. However, the transition region {$0.1 < Q^{2} < $ 1 GeV$^{2}$} remains beyond the capabilities of existing designs, creating the so-called ``$Q^{2}$ gap''~\cite{GDI}. Some possible approaches to bridge the $Q^{2}$ gap include lowering the electron beam energy, shifting the interaction-point position, and adding dedicated instruments.

Lowering the electron beam energy, $E$, helps because the minimum accessible $Q^{2}$ scales with $E^{2}$. While this is an elegant solution, it comes at the significant cost of reducing the center-of-mass energy of the collisions and, therefore, reducing the coverage at low values of $x$ (the minimum $x$ scales with $1/E$). Consequently, this approach does not provide access to the required kinematic region for gluon-saturation studies.

The second solution of shifting the interaction point was employed at HERA but is completely impractical at the EIC due to the presence of a beam crossing angle.

The third solution was implemented at HERA with the ZEUS Beam Pipe Calorimeter (BPC)~\cite{Surrow:1998su} and the H1 Very Low $Q^2$ spectrometer (VLQ)~\cite{Stellberger_2003}.  These detectors provided coverage down to $Q^{2}=0.11$ GeV$^{2}$ and $Q^{2}=0.08$ GeV$^{2}$, respectively.

The third option is the only viable one and is the focus of our work. In this study, we describe the design of a detector that will cover a few degrees from the electron-beam direction. We refer to this detector as the Few-Degree Calorimeter (FDC), which will cover the range of $-4.6 < \eta < -3.6$.

\begin{figure}[h!]
    \includegraphics[width=0.495\textwidth]{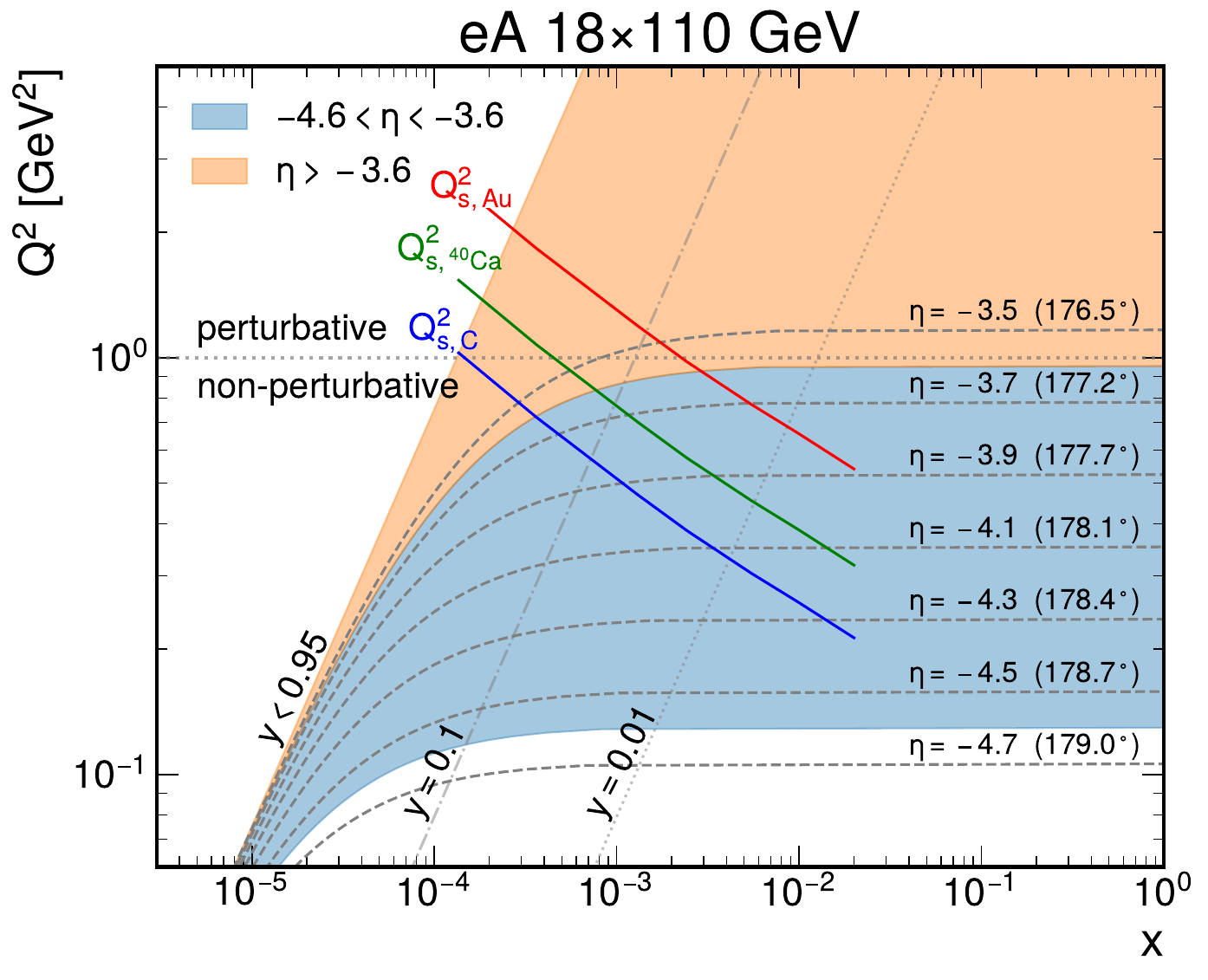}
    \caption{Coverage in the $x$ vs. $Q^2$ phase-space for EIC top-energy $eA$ collisions with the nominal range of the FDC (blue) spanning $-4.6<\eta<-3.6$, and the crystal ECAL (orange) spanning $\eta>-3.6$. The colored curves indicate the expected saturation scales, $Q_s(x)$, for various nuclei~\cite{AbdulKhalek:2021gbh,PhysRevLett.100.022303}.}
    \label{fig:x_Q2}
\end{figure}

Figure~\ref{fig:x_Q2} illustrates the coverage in the $x$ vs.~$Q^2$ phase-space within the nominal range of the FDC for the top-energy settings for $eA$ collisions. The FDC has the potential to significantly extend acceptance into a key kinematic phase-space and enable the study of the predicted gluon-saturation transition in inclusive DIS~\cite{PhysRevLett.86.596}, as well as many other observables such as exclusive vector-meson production~\cite{H1:2005dtp,ZEUS:2007iet}.

\section{Design Constraints and Requirements}
\label{sec:requirements}
\subsection{Location and Acceptance}
The primary challenge in measuring electron scattering at small angles lies in effectively instrumenting the region near the beampipe while minimizing the material in front of the detector. 

\begin{figure}[h]
    \centering
    \includegraphics[width=0.495\textwidth]{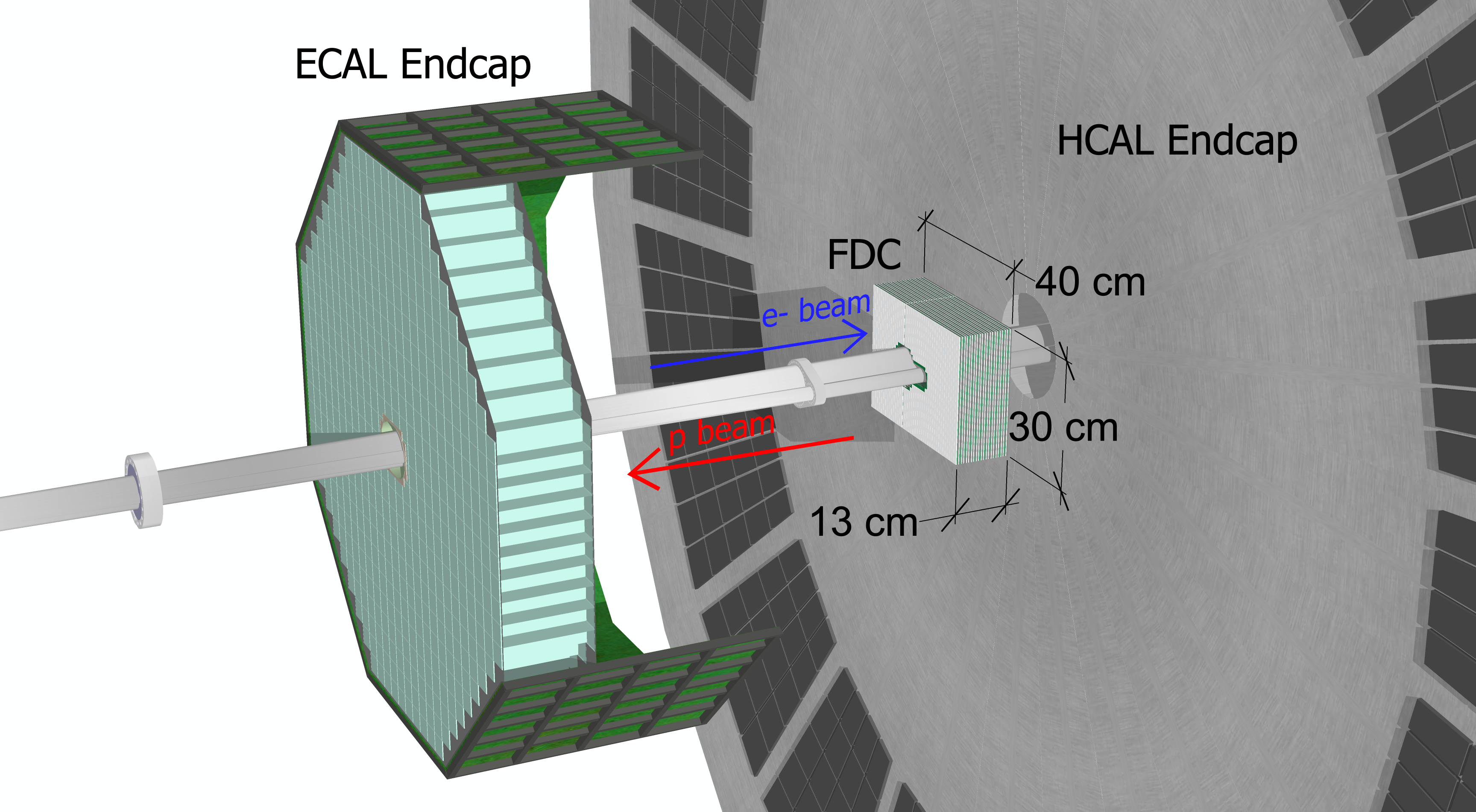}
     \includegraphics[width=0.495\textwidth]{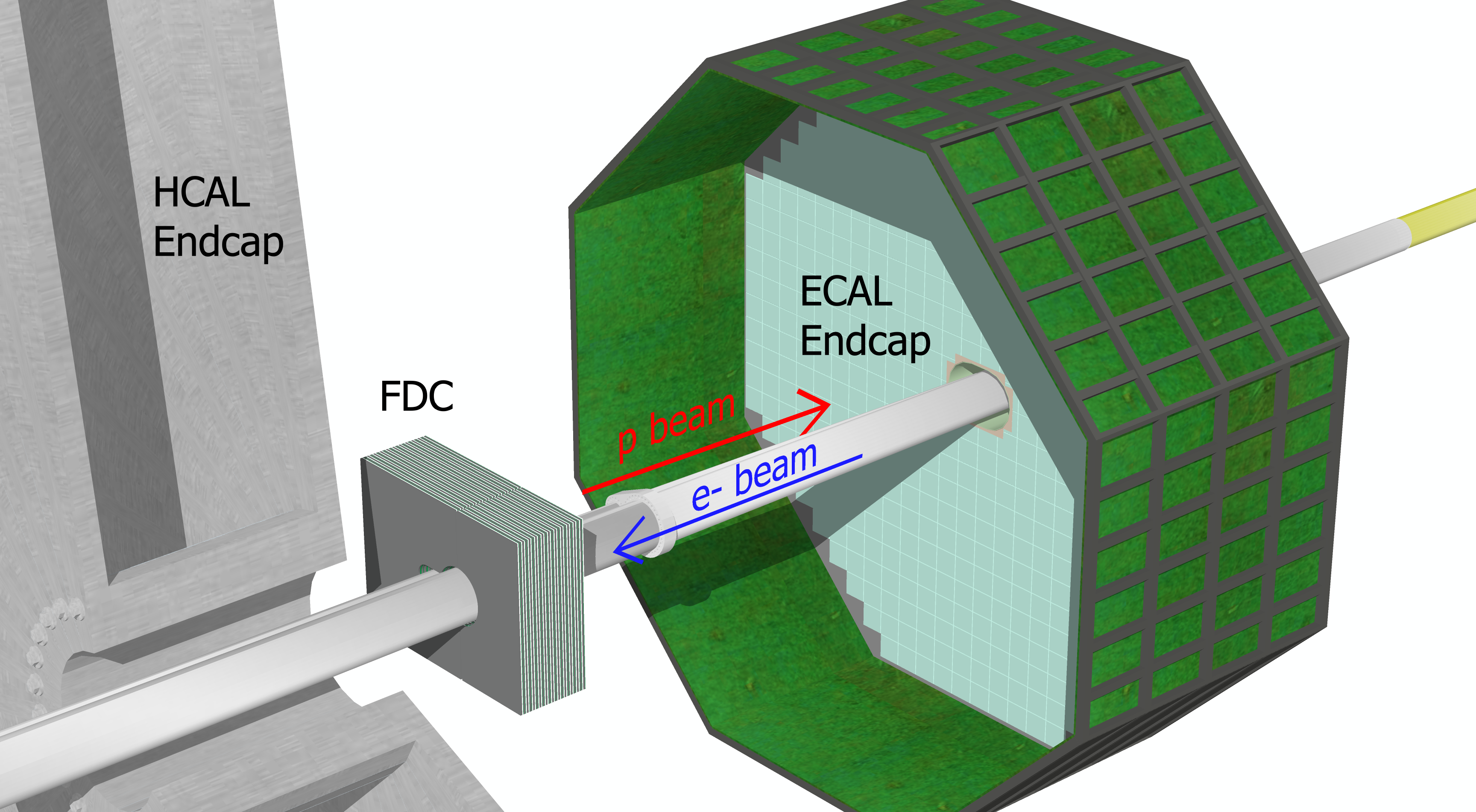}

    \caption{A potential detector layout showcasing the FDC, crystal ECAL, and endcap HCAL. The upper panel provides an upstream view, with the electron beam moving from left to right, while the bottom panel presents a downstream view. The \textsc{Sketchup} models for ECAL and HCAL can be found in Ref.~\cite{managerie}. The support structures to the floor are not shown for clarity.
}
    \label{fig:overview}
\end{figure}

One possibility is to position the FDC behind the crystal ECAL and in front of the backward HCAL, as illustrated in Fig.~\ref{fig:overview}. In the current ePIC design~\cite{managerie}, a potential location is at $z=-307$ cm, which would leave space for a compact calorimeter and about 10 cm gap before the HCAL. 

\begin{figure}[h!]
    \centering
    \includegraphics[width=0.495\textwidth]{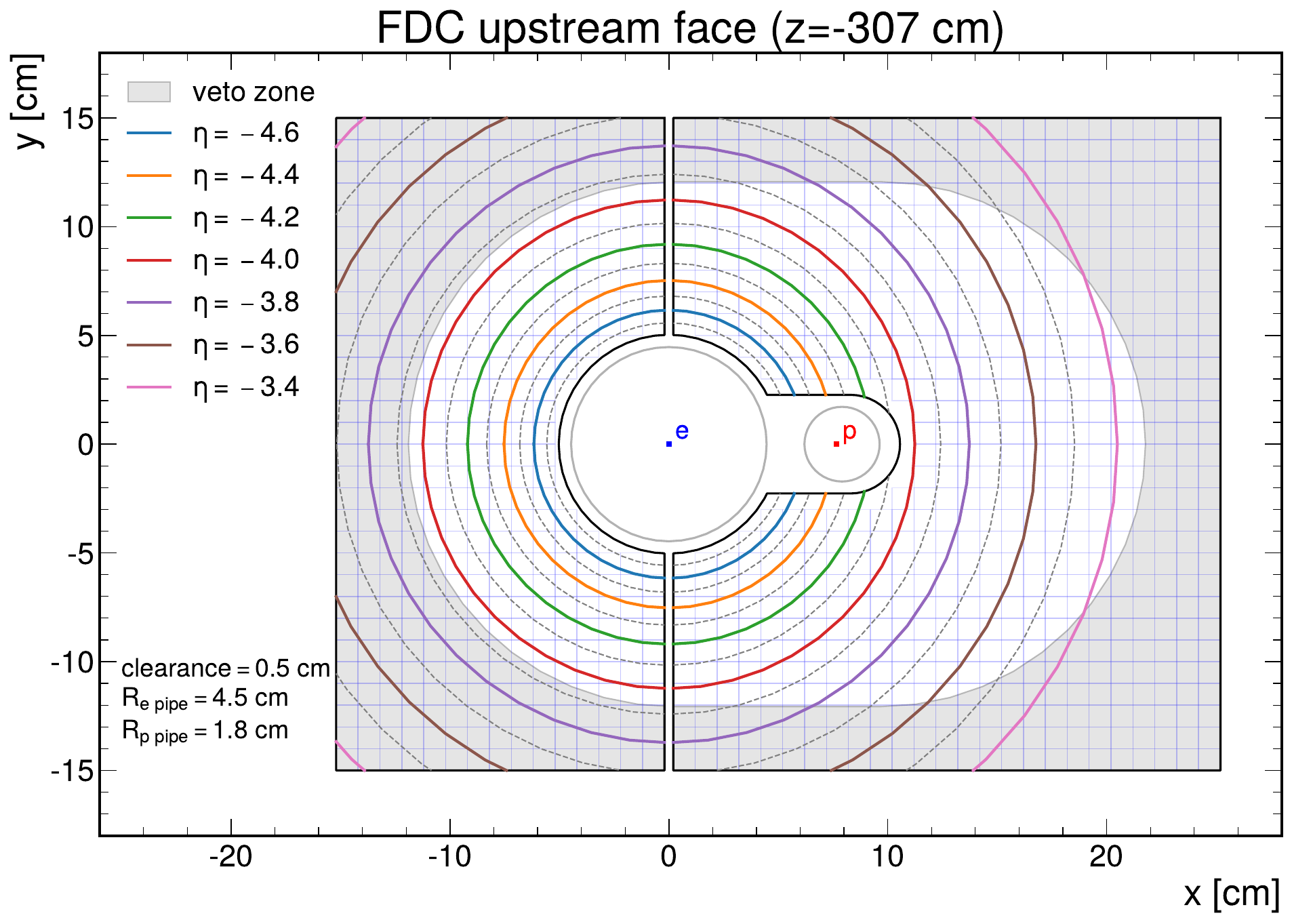}
    \caption{Transverse view of the FDC, assuming a location at $z=-307$ cm, with rings of constant $\eta$ superimposed. The electron and hadron beampipes are represented by their transverse cut, along with a $10\times10$ mm$^{2}$ grid for reference. The non-shaded area is a projection from the ECAL hole.}
    \label{fig:eta_rings}
\end{figure}

Figure~\ref{fig:eta_rings} shows that at $z=-307$ cm, the electron beampipe for IP6 has a radius of 4.5 cm, while the hadron beampipe has a radius of 1.8 cm and is shifted 8.3 cm in the $x$ direction. Assuming a 5 mm clearance to the beampipe, similar to the ZEUS BPC~\cite{Surrow:1998su}, a calorimeter with an outer perimeter of $30\times40$ cm$^{2}$ could cover the region $-4.6<\eta<-3.6$ with non-uniform azimuthal coverage. 

The shaded region on the FDC in Fig.~\ref{fig:eta_rings} represents the area where electrons would encounter part of the crystal ECAL or its support structure before reaching the FDC\footnote{We approximate the path of the electron as a straight line and assume that the flat cables servicing the ECAL SiPMs can be arranged to avoid the hole area.}. 
This region, which is a few cm wide, would serve as a veto area.

\begin{figure}[h!]
    \centering
    \includegraphics[width=0.495\textwidth]{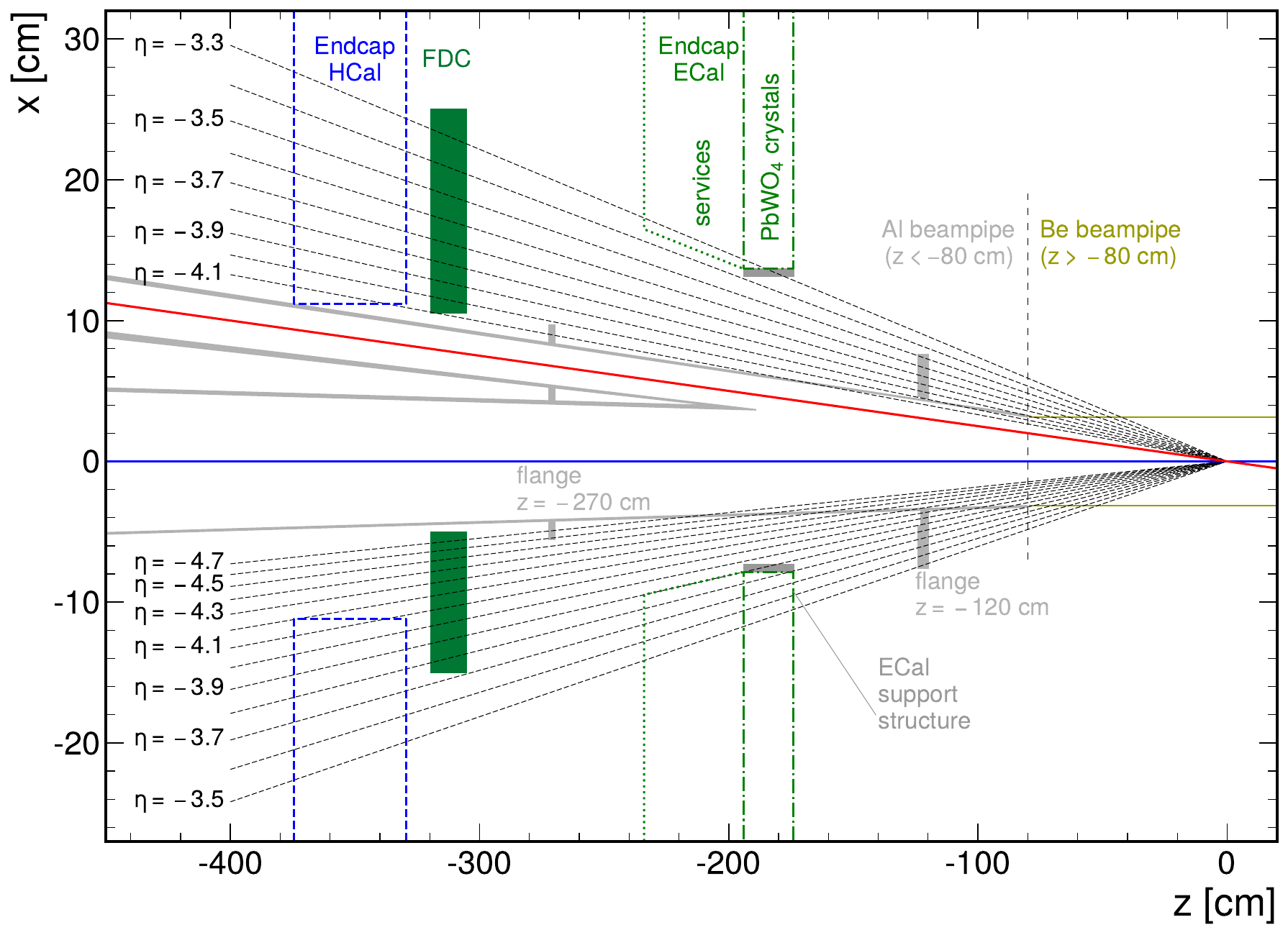}\\
    \includegraphics[width=0.495\textwidth]{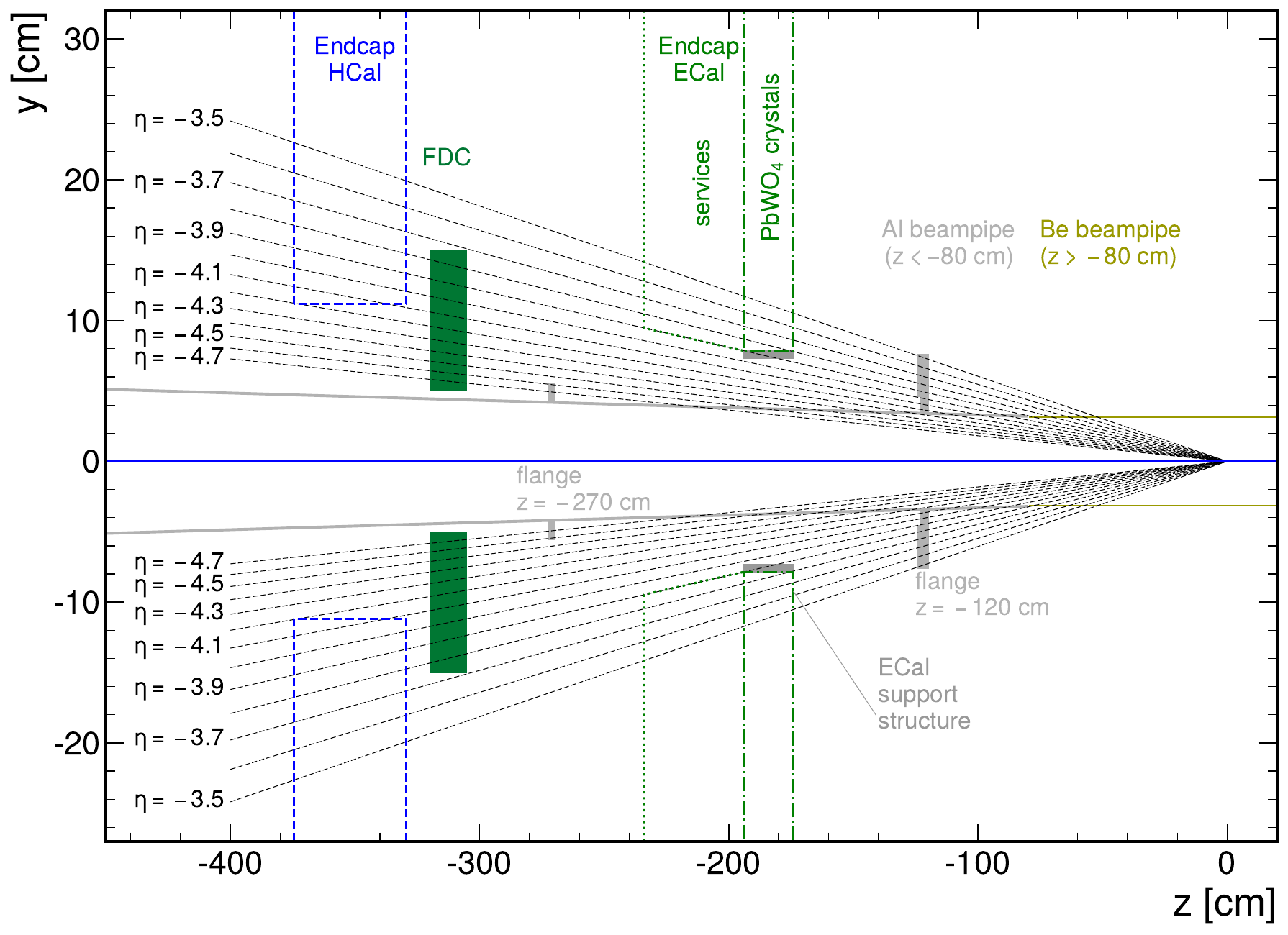}

    \caption{A potential location for the FDC is behind the backward ECAL in the current ePIC design. This diagram illustrates the dimensions and locations of ECAL and HCAL as per Ref.~\cite{managerie}; the flanges and clearance dimensions as per Ref.~\cite{ELKEprivate}, and ECAL support structure as per Ref.~\cite{MechanicalDesign}. The blue and red lines indicate the electron and hadron beampipes, respectively.}
    \label{fig:FDCposition}
\end{figure}

\newpage
Figure~\ref{fig:FDCposition} shows projections of the possible detector layout, including both the $yz$ and $xz$ views. The ECAL hole is currently assumed to have a height of 14.7 cm and a width of 20.5 cm, taking into account the current version of the ``micro-flange'' (with a cam shape that is 15.2 cm wide and 11.1 cm tall) and a required clearance of 1.8 cm, 3.6 cm, 1.8 cm, and 1.7 cm between the flange and the ECAL's inner support structure on the top, right, bottom, and left sides when looking downstream~\cite{ELKEprivate}. 

\subsection{Dead Material in Front of FDC}
The main challenge faced by a detector located in a high pseudorapidity range is that particles can encounter a significant amount of material as they graze the beampipe walls. While converted electrons and photons can be identified through shower-shape information, reducing the amount of material helps by improving efficiency and minimizing background. 

\begin{figure}[h!]
\centering
\includegraphics[width=0.495\textwidth]{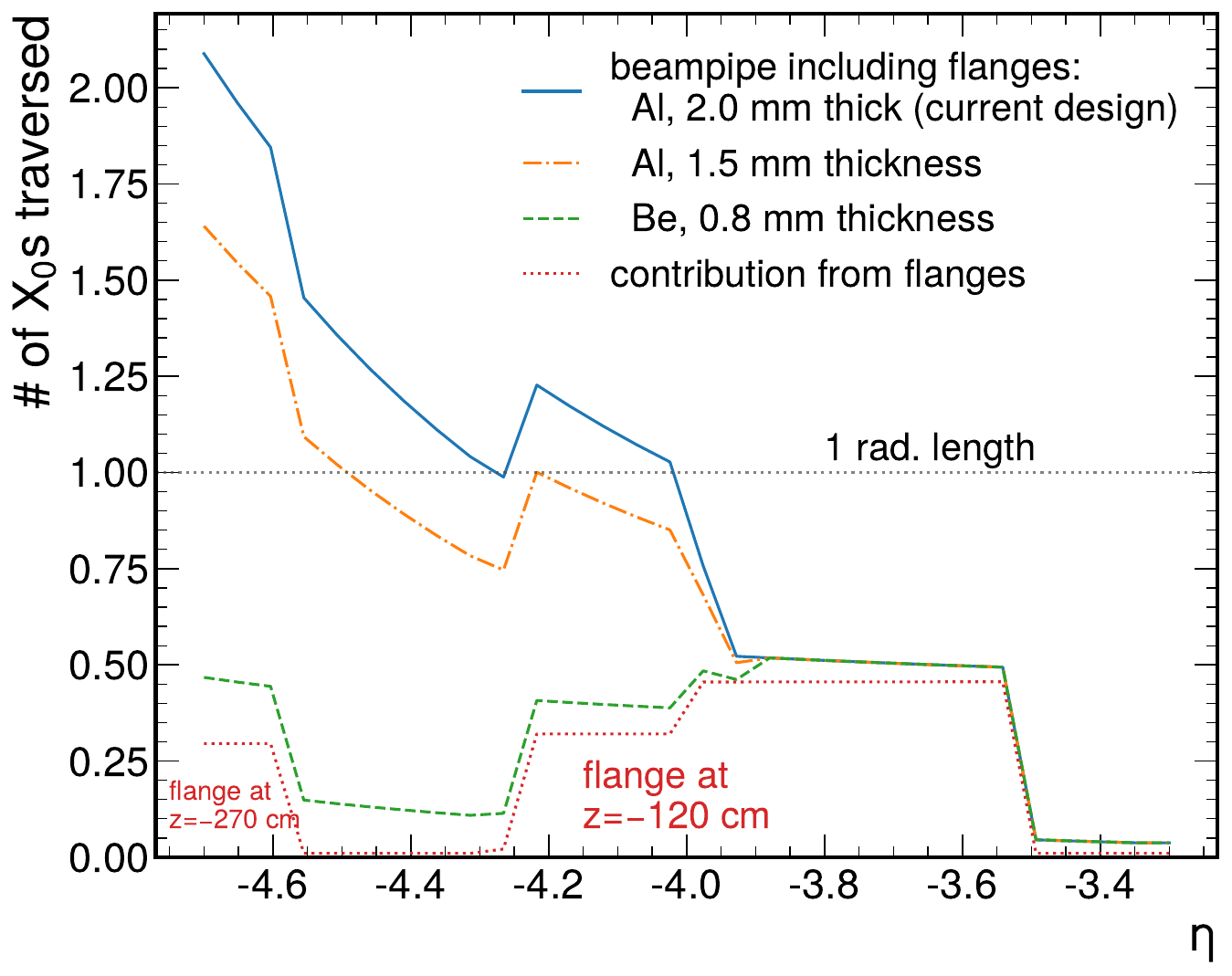}
\caption{Radiation lengths traversed in the electron-beampipe material at $\phi = 180^{\circ}$ as a function of $\eta$. It includes the IP6 beampipe model (solid blue), a modified version with  thinner aluminum (orange dot-dashed), and another scenario with beryllium (green dashed). The contribution from the flanges is represented by the red dotted curve. }
\label{fig:rad_lengths}
\end{figure}

Figure~\ref{fig:rad_lengths} illustrates the number of radiation lengths of beampipe material encountered by electrons before reaching the FDC, as a function of $\eta$. Within most of the FDC acceptance, the total number of radiation lengths ranges from 0.5 to 1.2. Approximately half a radiation length is contributed by the flange at $z=-120$ cm within the range of $-4.2 < \eta < -3.5$. The significant increase in the total number of $X_0$ traversed at around $\eta=-4.0$ is attributed to the use of aluminum instead of beryllium in the beampipe for $z < -80$ cm.

 The HERA experiments successfully addressed this challenge by implementing a beampipe ``exit window'' made of thin aluminum, which reduced the total dead material to less than 1 $X_{0}$~\cite{Stellberger_2003}. Alternatively, one could use a beryllium section. 

Figure~\ref{fig:rad_lengths} illustrates that incorporating a beryllium section, within the range of $-205 < z < -80$ cm, would effectively decrease the overall material budget in the $-4.7 < \eta < -4.1$ region to below 0.5 $X_{0}$. As an alternative, implementing a 1.5 mm aluminum layer would yield a reduction to less than 1 $X_{0}$ for $\eta > -4.5$.

Another effective method for mitigating the impact of dead material is to use shower shapes to tag converted electrons or photons that originated further upstream in the beampipe or flanges~\cite{Surrow:1998su}. 

\subsection{Acceptance Limit}
\begin{figure}[h!]
    \centering
    \includegraphics[width=0.495\textwidth]{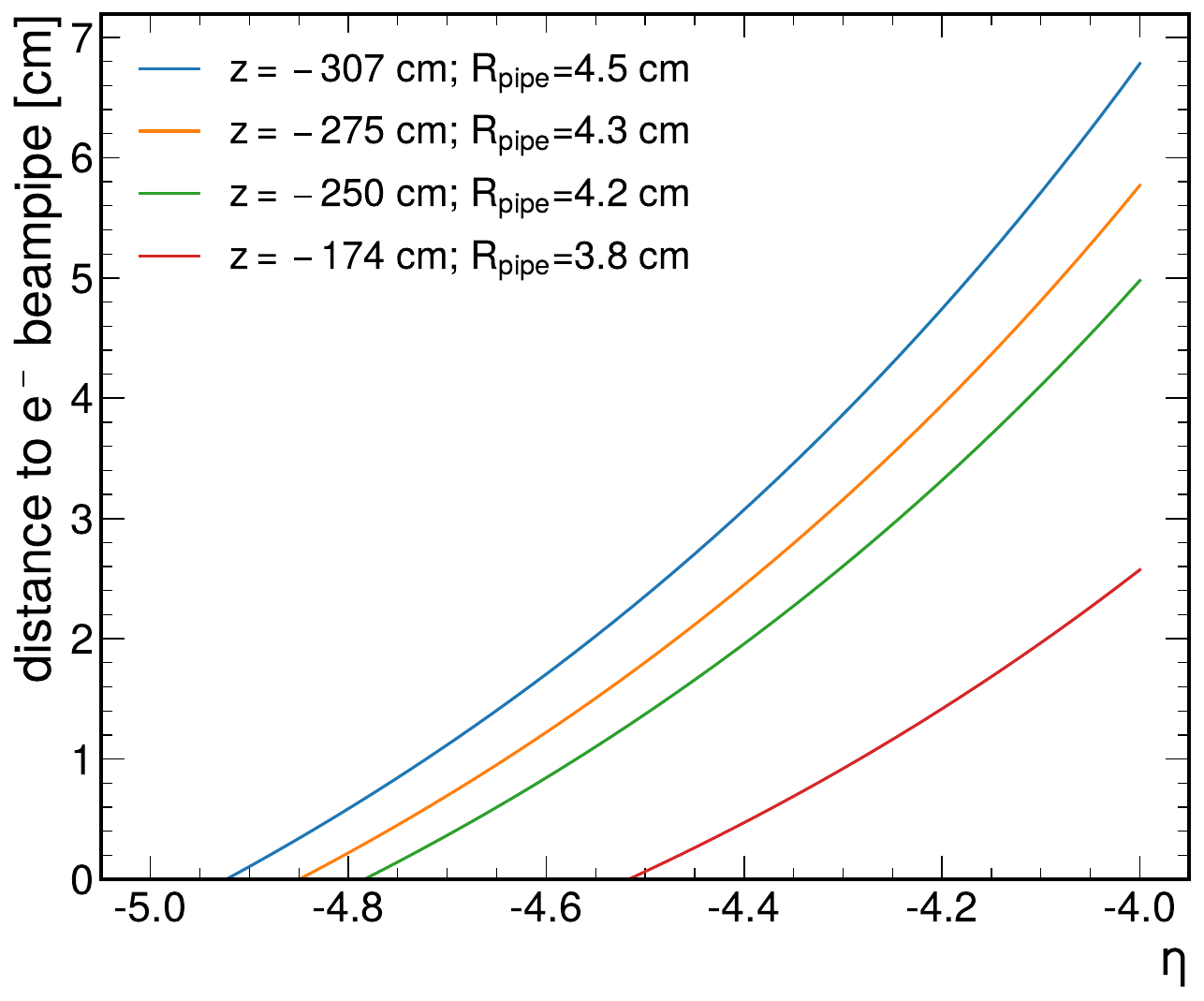}
    \caption{Distance from the electron beampipe as a function of $\eta$ at various $z$ locations, including the proposed FDC position (blue) and the location of the ECAL endcap (red).}
    \label{fig:distance_from_beampipe}
\end{figure}

To enable accurate FDC measurements at small angles, it is crucial to minimize energy leakage into the beampipe. Figure~\ref{fig:distance_from_beampipe} shows the distance to the beampipe surface as a function of $\eta$.
The target value of $\eta=-4.6$ corresponds to about 18 mm from the electron beampipe at $z=-307$ cm. Assuming a 5 mm clearance, this leaves 13 mm between the edge of the detector and $\eta=-4.6$. For reference, the ZEUS BPC measured 95$\%$ of the energy from 5 GeV electrons at 8 mm from the detector's edge in test beams~\cite{Surrow:1998su}.

\subsection{Energy Range}
\begin{figure}[h!]
    \centering
    \includegraphics[width=0.495\textwidth]{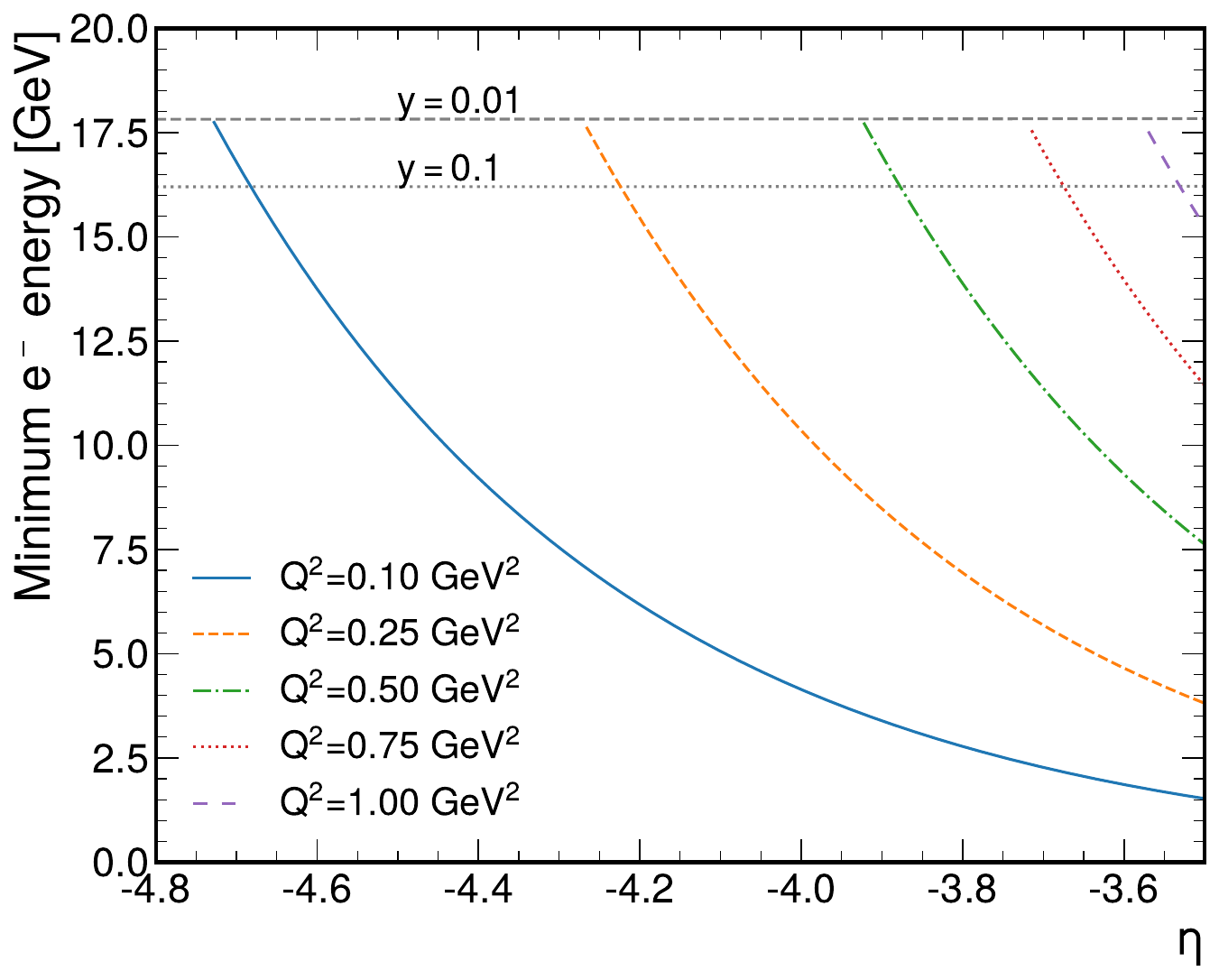}
    \caption{Minimum energy of electrons for a given $Q^2$ as a function of $\eta$, for an electron-beam energy of 18 GeV (independent of hadron or ion-beam energy). }
    \label{fig:MinEnergy}
\end{figure}

The main objective of the FDC is to identify and measure the energy and angle of electrons in the $0.1<Q^{2}<1.0$ GeV$^{2}$ range. Figure~\ref{fig:MinEnergy} illustrates the minimum electron energy as a function of $\eta$ for various $Q^2$ values. The minimum energy required for $-4.6<\eta<-3.6$ falls within the range of 2--13 GeV, whereas the maximum is the beam energy, 18 GeV. Thus, the target energy range is 2--18 GeV.

\subsection{Background Rejection}
\label{sec:bkgrejection}
\begin{figure}[h!]
    \centering
    \includegraphics[width=0.495\textwidth]{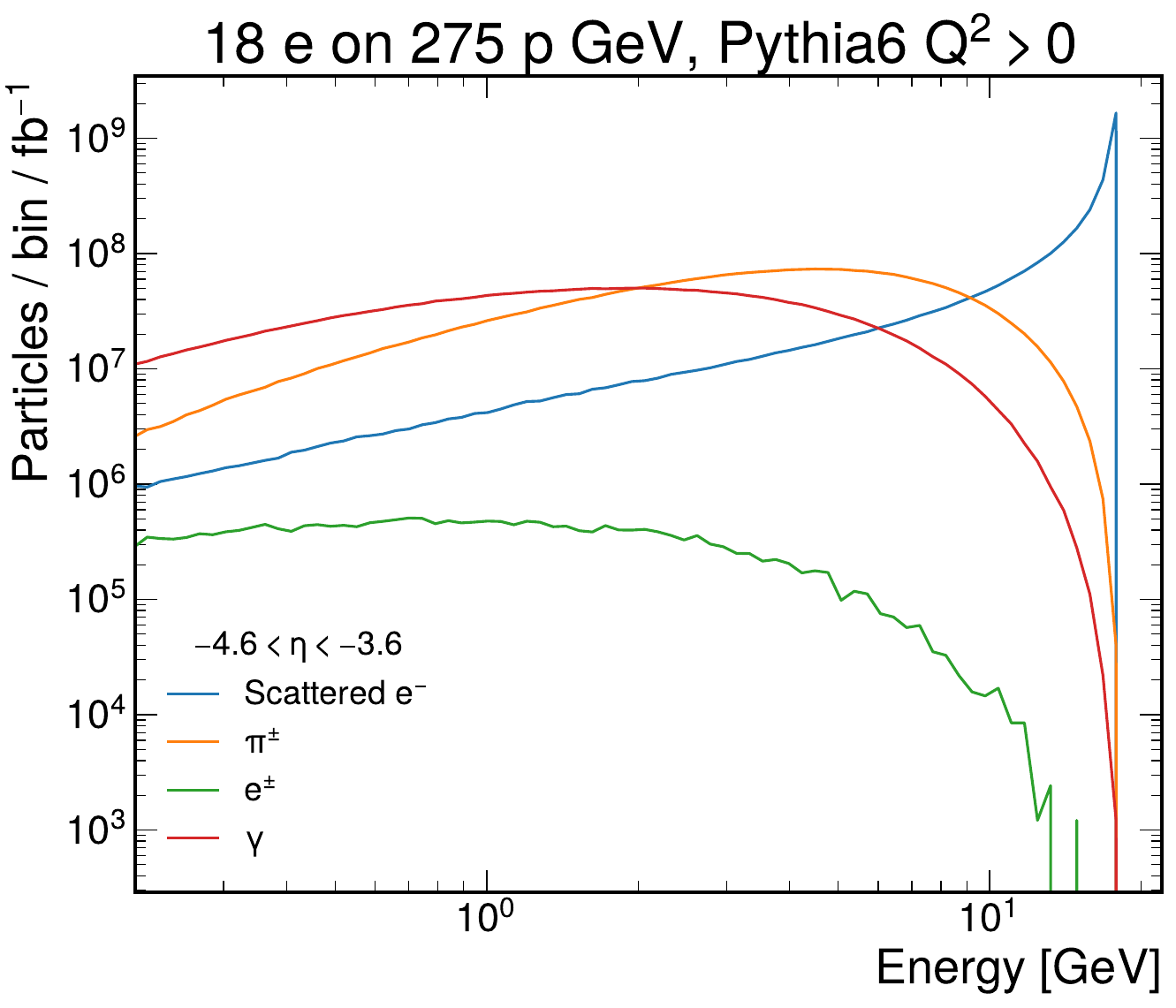}
    \caption{Particle spectra per fb$^{-1}$ of integrated luminosity, estimated using \textsc{Pythia6} with no $Q^2$ cut.}
    \label{fig:bkg}
\end{figure}

The main background for inclusive DIS measurements originates from events with small $Q^{2}$, where the scattered electron is not detected, but an electron candidate is observed in the FDC. Figure~\ref{fig:bkg} shows the expected particle spectra obtained using \textsc{Pythia6} to simulate $ep$ scattering without a $Q^2$ cut and without considering detector effects. In the absence of charge tagging, both electrons and positrons from semi-leptonic decays contribute to the background. Similarly, the charged-pion background includes both charges. The photon background primarily arises from neutral-pion decays.

 We estimate the background rejection power of approaches employed at HERA~\cite{ZEUS:1997etp, Surrow:1998su} and the EIC YR. Specifically, we use the far-backward detectors as a veto for photoproduction and select events based on their energy-momentum imbalance\footnote{The $E-p_{z}$ distribution peaks at twice the electron-beam energy when the scattered electron is correctly identified.} with a loose selection of $E - p_{z}>18$ GeV~\cite{AbdulKhalek:2021gbh}.

\begin{figure}[h!]
    \centering
        \includegraphics[width=0.495\textwidth]{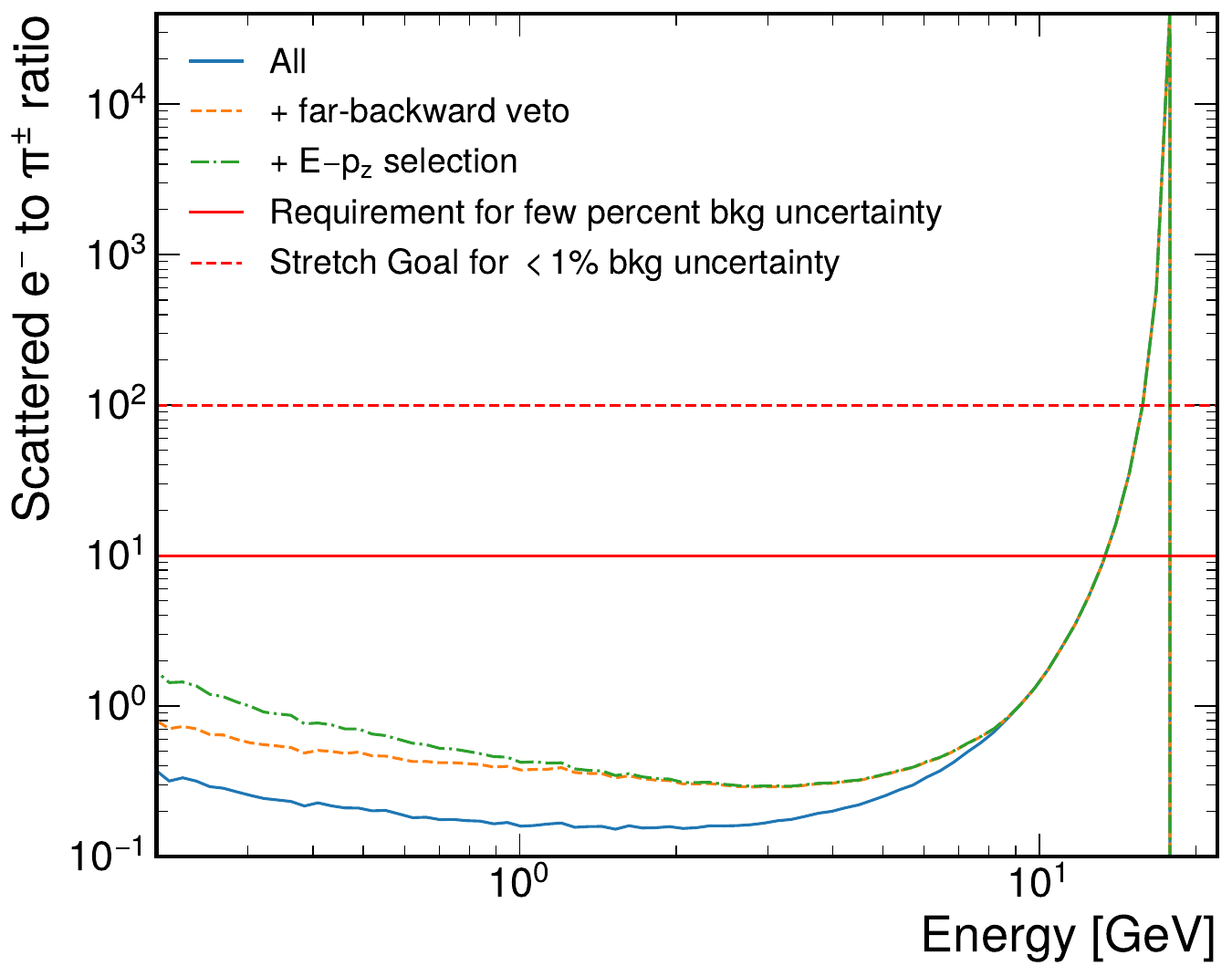}
    \caption{Ratio of scattered electrons to charged pions as a function of energy.}
    \label{fig:bkgratio}
\end{figure}

 Figure~\ref{fig:bkgratio} illustrates the impact of these cuts on the $e/\pi$ ratio. The far-backward veto has a modest effect due to its small acceptance~\cite{AbdulKhalek:2021gbh}. Similarly, the $E - p_{z}$ cut has a modest impact, since for background events the scattered electron has low energy. Overall, the resulting $e/\pi$ ratio is about $e/\pi\approx$ 0.4 between 1--6 GeV, $e/\pi\approx$ 1 around 10 GeV, and increases rapidly at higher energies. Similarly, the resulting $e/\gamma$ ratio ranges from 0.1 to 1 for $E<6$ GeV and increases at higher energies. The background of positrons and electrons from semi-leptonic decays reaches 10\% at 1 GeV, decreasing to less than 1\% at 10 GeV. Since there is no magnetic field near the FDC, this background cannot be subtracted using reverse-field runs. Instead, Monte-Carlo studies would be employed to estimate and subtract it. Alternatively, electron-isolation criteria might reduce this background further. 

The FDC measurements should aim for a purity greater than 90\% to achieve background uncertainties at the few percent level. A stretch goal of 99\% purity would result in a negligible uncertainty compared to the expected luminosity uncertainty of 1\%~\cite{AbdulKhalek:2021gbh}. Hence, the necessary rejection power falls within the range of 10-25 for $\pi^{\pm}$ and 10-100 for $\gamma$, with a factor of 10 higher for the stretch goal.

The standalone FDC's $\pi^{\pm}$ rejection power will rely on its shower-shape capabilities. Longitudinal segmentation can play a crucial role in discriminating hadrons, as they are more likely to interact deeper within the detector, while electrons tend to exhibit showers starting primarily in the initial layers. Moreover, transverse segmentation also helps, as electrons typically produce narrower and more regular showers compared to hadronic ones. This approach is expected to work well above a few GeV. 

\begin{figure}[h!]
    \centering
    \includegraphics[width=0.495\textwidth]{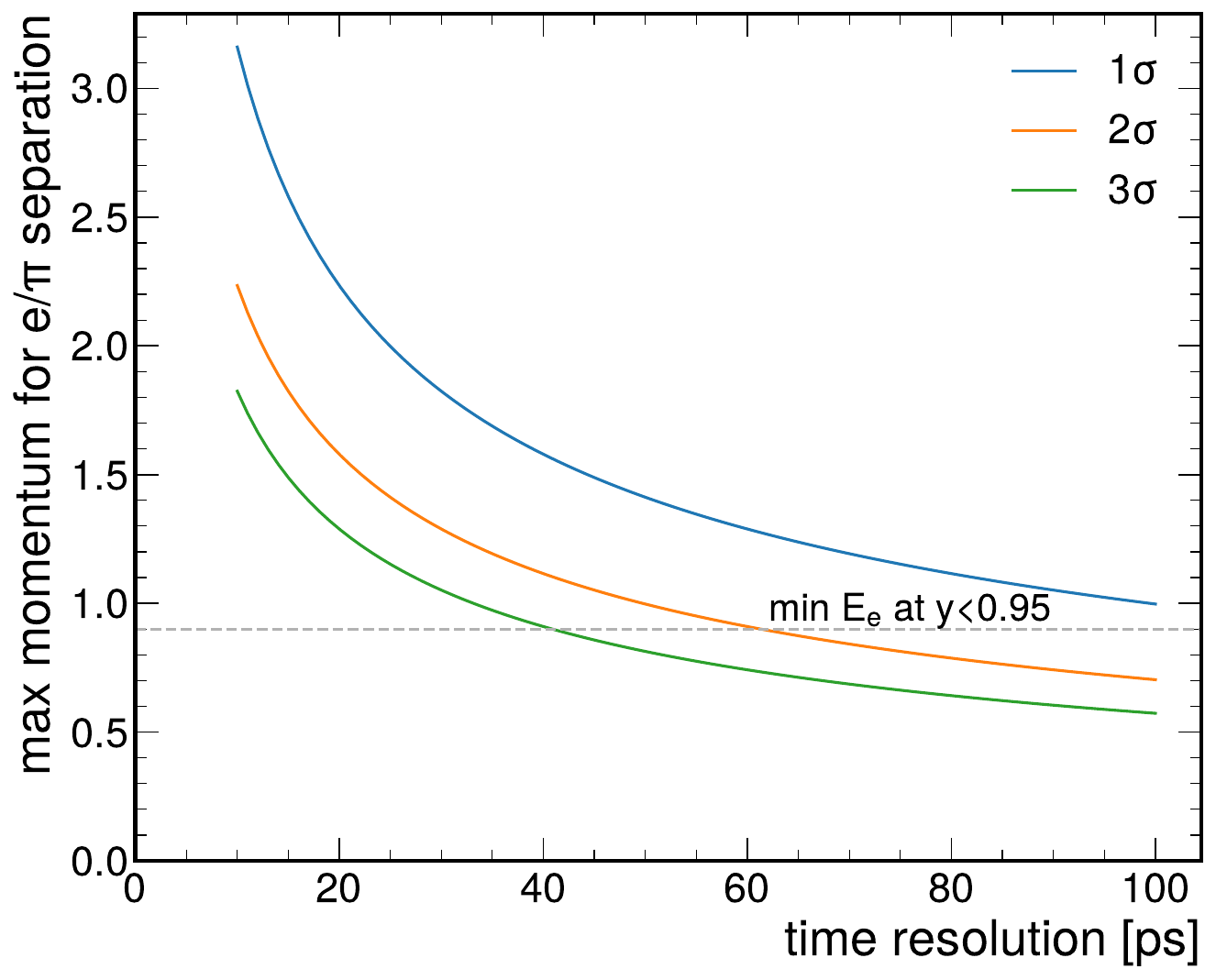}
    \caption{Required time resolution to discriminate between electrons and charged pions using TOF as a function of the particle's momentum. These requirements are based on a location at $z=-307$ cm.}
    \label{fig:time_requirement}
\end{figure}

Additional background rejection can be achieved through auxiliary systems, such as a scintillator layer for tagging MIPs to reject unconverted photons, or a timing layer to reject low-energy hadrons. Figure~\ref{fig:time_requirement} demonstrates the potential of TOF and illustrates that a time resolution of about 50~ps would be necessary to achieve a 2$\sigma$ $e/\pi$ separation below 1~GeV.

\section{Design}
\label{sec:design}

The FDC design draws inspiration from the ZEUS BPC~\cite{Surrow:1998su} and H1 VLQ~\cite{Stellberger_2003} calorimeters, incorporating modern enhancements in optical readout and photosensors. It leverages recent advancements in high-granularity calorimetry~\cite{Sefkow:2015hna}, led by the CALICE collaboration. These developments offer the potential for substantial improvements in granularity at a reasonable cost for a small detector such as the FDC.

The CALICE collaboration has tested a scintillator-tungsten ECAL with wavelength-shifting fibers coupled to SiPMs~\cite{CALICE:2017sis}. The tested strips had widths of 10 mm. By applying a ``split-strip algorithm''~\cite{KOTERA2015158}, this prototype can achieve an effective granularity close to 10$\times$10 mm$^{2}$. The test-beam data resulted in an energy resolution of 12$\%/\sqrt{E}\oplus1.2\%$. More recently, this design has been superseded by a ``SiPM-on-tile'' approach, where the SiPM is air-coupled to the scintillator strip~\cite{Dong:2018hvs,Niu:2020iln}~\footnote{This approach is also used in the ePIC calorimeter insert~\cite{Arratia:2022quz,Arratia:2023rdo}.}.

Table~\ref{tab:summary} summarizes our target design in comparison with the ZEUS BPC and H1 VLQ designs. 
\begin{table}[h!]
   \centering
   \caption{Summary description of our proposed EIC FDC, the ZEUS BPC~\cite{Surrow:1998su}, and the H1 VLQ~\cite{Stellberger_2003}.} 
   \begin{tabular*}{.50\textwidth}{@{\extracolsep{\fill}}llll@{}}
    \hline
       & EIC FDC &  ZEUS BPC & H1 VLQ \\
      \hline
        Depth & 20 $\mathrm{X}_{0}$ & 24 $\mathrm{X}_{0}$ & 16.7 $\mathrm{X}_{0}$ \\
        W/Sc thickness & 3.5/2 mm & 3.5/2.6 mm  & 2.5/3 mm  \\
        Moliere Radius & 15 mm & 13 mm & 15 mm\\
        Optical readout & SiPM-on-tile & WLS bar& WLS bar
        \\
        & & + PMT & +PIN \\ 
        Trans. granularity & 10$\times$50 mm$^{2}$ & 7.9$\times$150 mm$^{2}$ & 5$\times$120 mm$^{2}$\\
        Long. granularity & every strip & none & none \\
        Channels & 4500 & 31 & 336  \\ 
        Readout & HGROC & FADC/TDC & ASIC  \\ 
        Position res. & 3.6 mm$/\sqrt{E}$ &  2.2 mm$/\sqrt{E}$ & 2 mm$/\sqrt{E}$\\
        Energy res. & $\frac{17\%}{\sqrt{E}}\oplus2\%$ & $\frac{17\%}{\sqrt{E}}\oplus2\%$ & $\frac{13\%}{\sqrt{E}}\oplus3\%$  \\
        Time resolution &   $<$50 ps & 400 ps & ---\\

\hline

  \end{tabular*}
   \label{tab:summary}
\end{table}

\begin{figure*}[h!]
    \centering
    \includegraphics[width=0.429\textwidth]{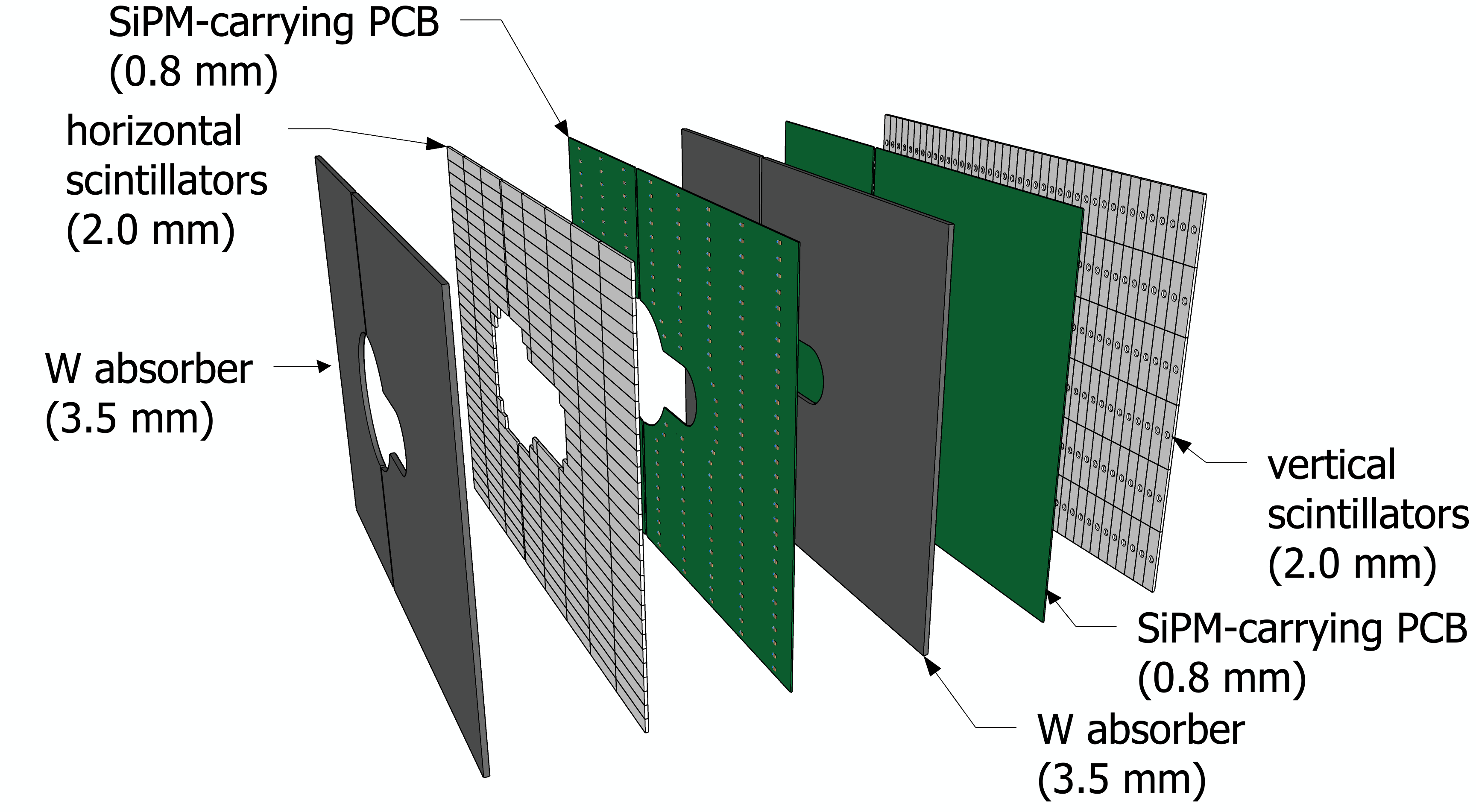}
        \includegraphics[width=0.249\textwidth, trim={25cm 0 25cm 0},clip]{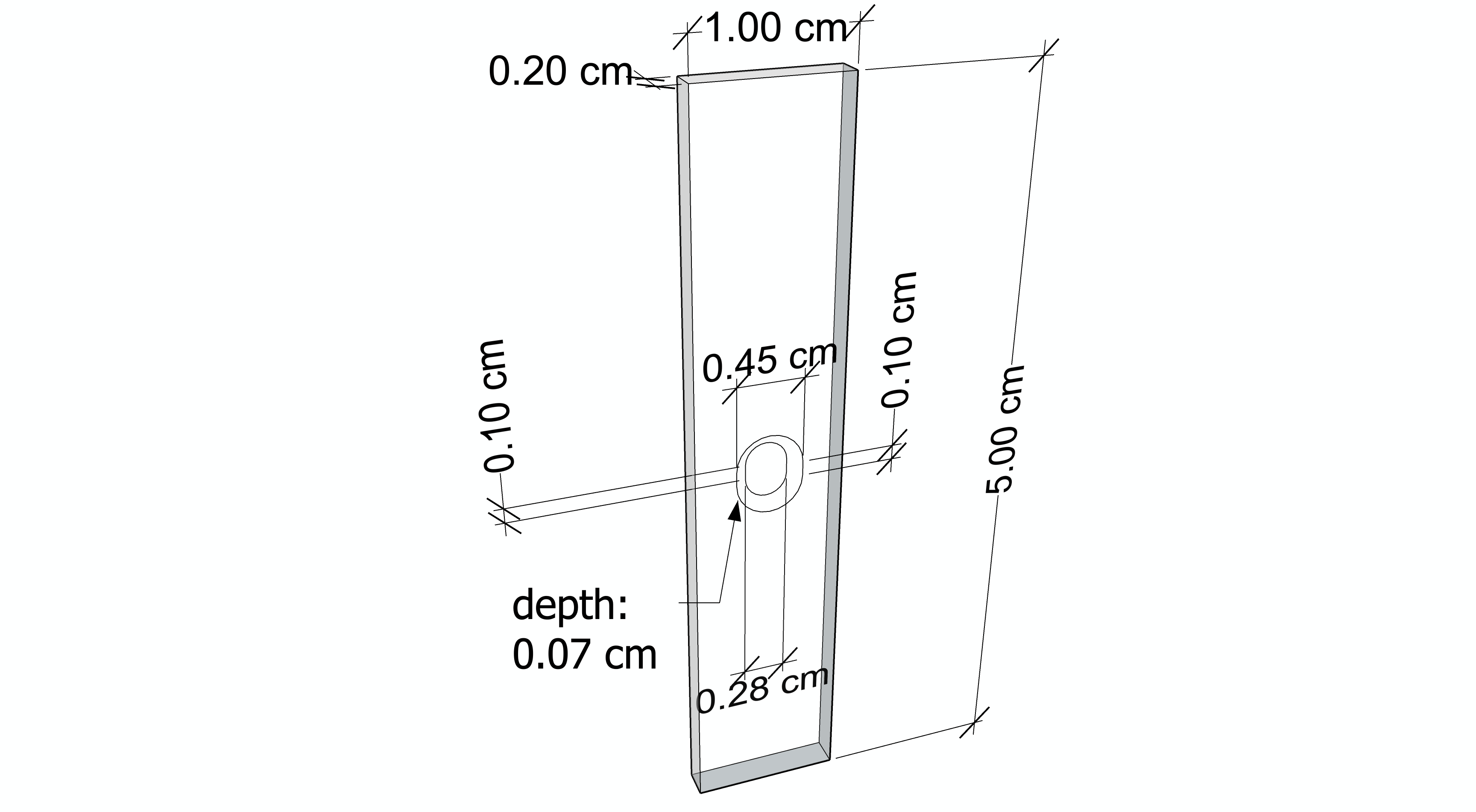}
            \includegraphics[width=0.312\textwidth, trim={15cm 0 15cm 0},clip]{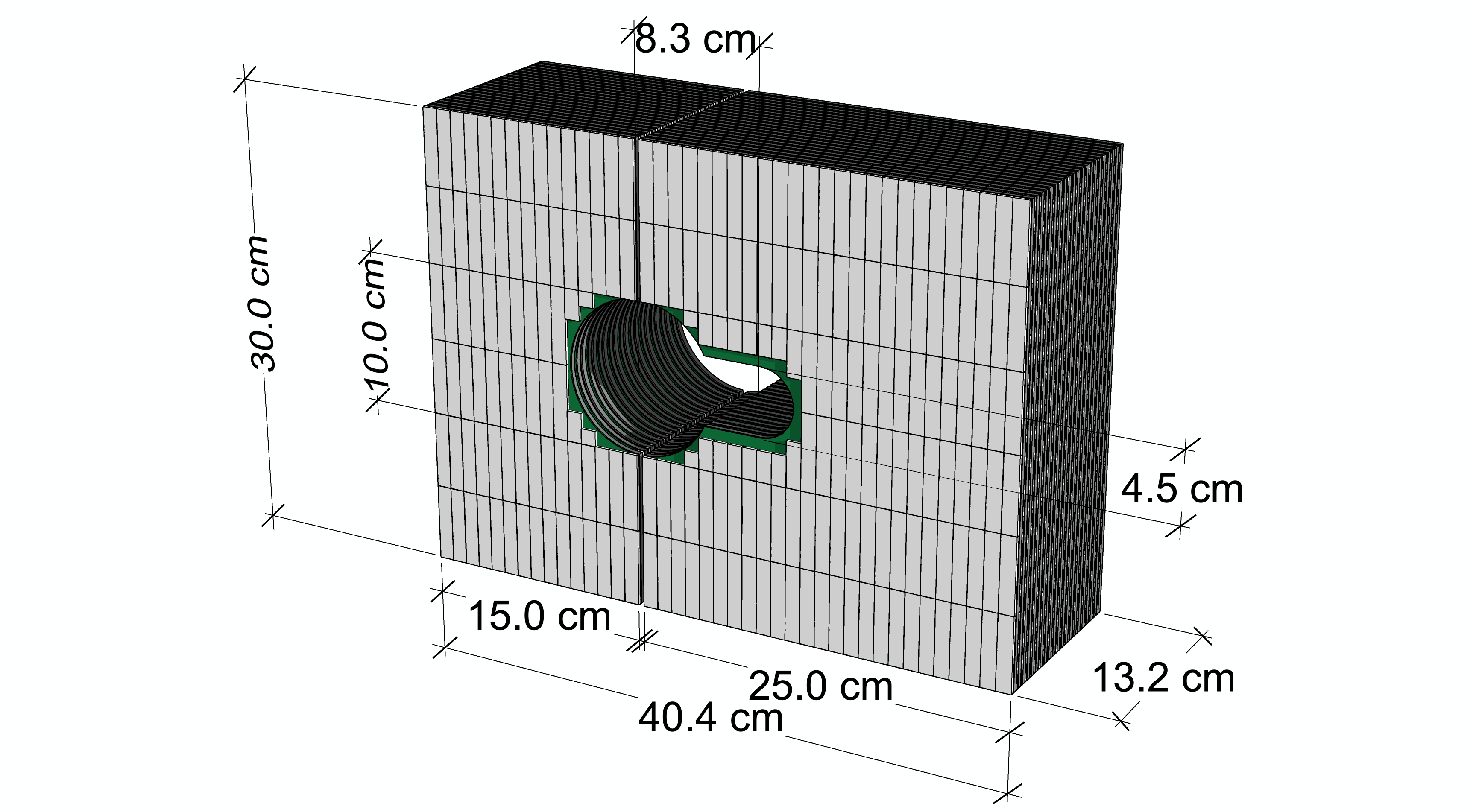}
    \caption{Top: exploded view of a double layer of the FDC. Middle: dimensions of a single scintillator strip. Right: Dimensions of the FDC.}
    \label{fig:explode_view}
\end{figure*}

Figure~\ref{fig:explode_view} shows the FDC design which includes alternating layers of vertical and horizontal scintillators that are wrapped in reflective foil and read out using SiPMs (HPK 14160-1315PS). The scintillator strips measure $50\times10\times2$ mm (length, width, thickness) and feature a dimple at the center for air-coupling with the SiPM.

Each tungsten layer is 3.5 mm (1 $X_{0}$, and 0.035$\lambda$). The total FDC consists of 20 layers, for a total of 20 $X_{0}$. Up to about $\eta=-4.1$, the detector has full coverage in $\phi$, whereas at larger negative $\eta$, the hole for the hadronic beampipe removes up to $60^\circ$ of acceptance.  

The FDC is divided into two parts, allowing one half to be retracted to the left and the other half to the right for maintenance purposes of the SiPM boards, particularly in the case of unexpected radiation loads. However, it is worth noting that SiPM annealing should not be necessary given that at the FDC location, the neutron flux is moderate and reaches approximately 10$^{9}$ 1-MeV equivalent neutrons per cm$^{2}$ for a year of running at top luminosity~\cite{Doses}.


\section{Simulation}
\label{sec:simulation}

We used the \textsc{DD4HEP} framework~\cite{Frank:2014zya} to run \textsc{Geant4}~\cite{GEANT4:2002zbu} simulations of electrons generated with a uniform azimuthal angle at various $\eta$ points. The simulation does not include any dead material, the effects of which we leave for future work. 
\subsection{Energy Resolution}
\begin{figure}[h!]
    \centering
    \includegraphics[width=0.465\textwidth]{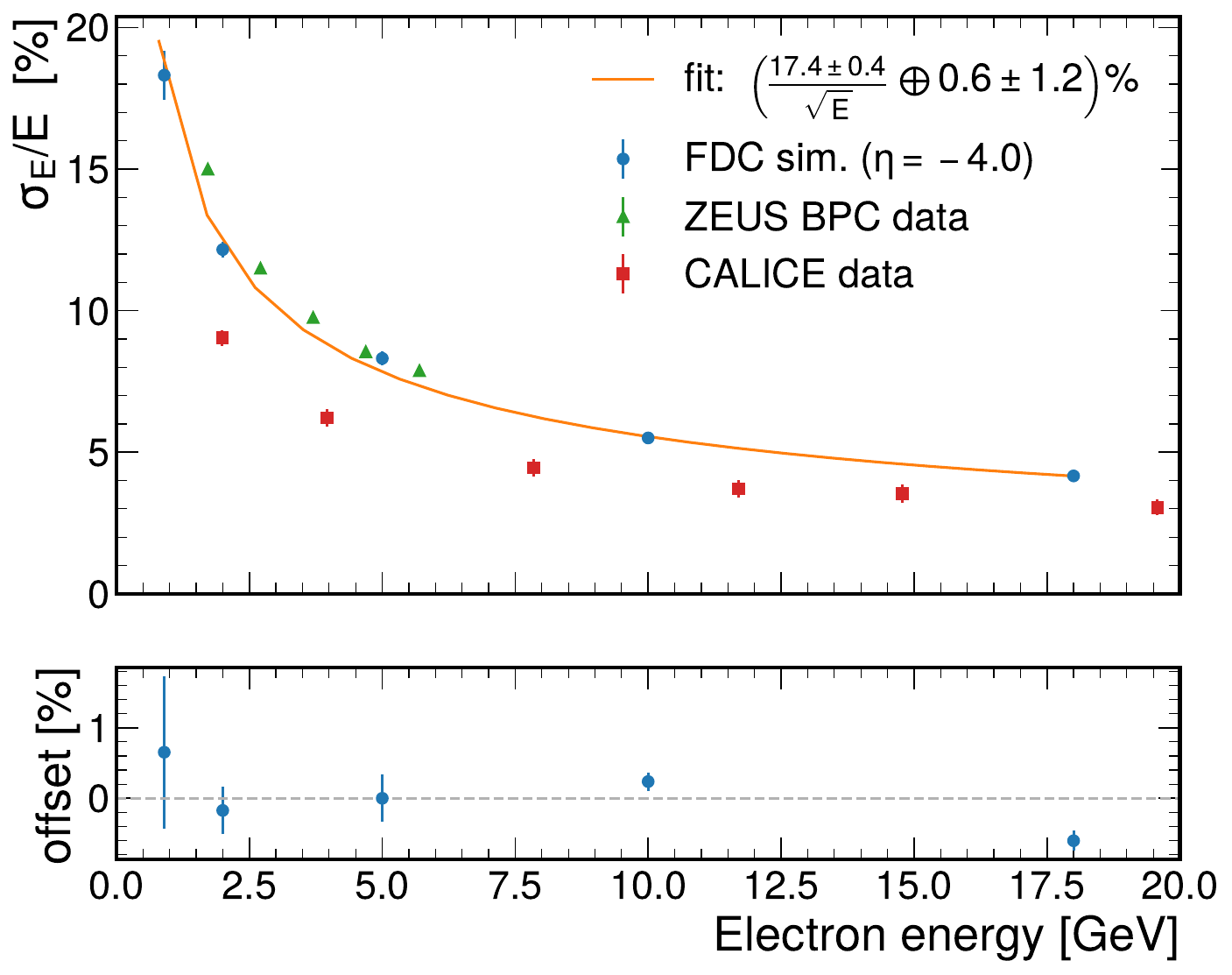}
    \caption{FDC energy resolution at $\eta=-4.0$ compared to ZEUS BPC~\cite{Surrow:1998su} (green triangles) and CALICE~\cite{KOTERA2015158} (red squares) data.}
    \label{fig:recon_E}
\end{figure}

Figure~\ref{fig:recon_E} shows the energy resolution, which can be parameterized as 17$\%/\sqrt{E} \oplus 2\%$, and is consistent with the ZEUS BPC data~\cite{Surrow:1998su}. We also compare it to CALICE data, which exhibits improved performance at the expense of a larger Moli\`ere radius. 

\begin{figure}[h!]
    \centering
\includegraphics[width=0.465\textwidth]{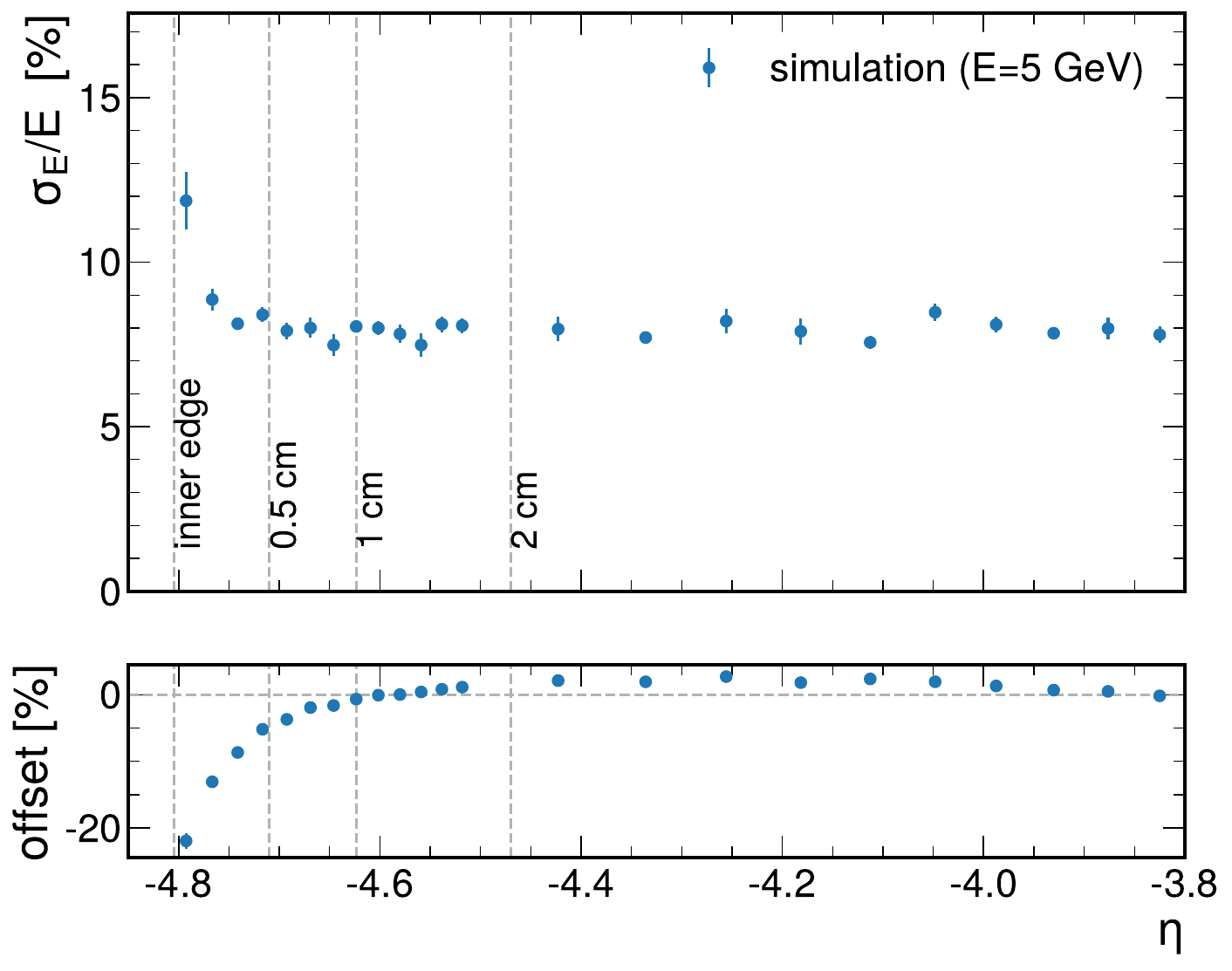}
    \caption{Dependence of the energy resolution on $\eta$. Vertical lines indicate different distances from the edge of the hole.}
    \label{fig:edge}
\end{figure}

Figure~\ref{fig:edge} shows the energy resolution and scale as a function of $\eta$. The performance remains relatively stable for $\eta>-$4.6.
 For particles that hit near the edge of the detector's upstream face ($\eta\approx-4.8$), the energy-scale offset is $-$20\%, and the resolution is about 12\%. One would expect that half the shower would be in the calorimeter (a loss of $50\%$), but because the hole is a cylinder, the center of the shower moves further away from the hole as it goes through the calorimeter.

\subsection{Position Resolution}
The polar-angle resolution is determined by considering the position resolution of the FDC and the resolution of the vertex position. The electron vertex position can be precisely determined by tracking other particles in the event using the main detectors, as was done in HERA~\cite{Surrow:1998su,Stellberger_2003}. Thus, only the FDC position resolution is relevant.

We reconstructed the $x$ and $y$ values following the method described in Ref.~\cite{Monteiro:1998bi}:
\begin{equation}
x=\frac{\sum\limits_{i\in v\,layers} w_{X,i} x_i}{\sum\limits_{i\in v\,layers} w_{X,i}}, 
y=\frac{\sum\limits_{i\in h\,layers} w_{Y,i} y_i}{\sum\limits_{i\in h\,layers} w_{Y,i}}
\end{equation}
where the weights $w_{X,i}$ and $w_{Y,i}$ are determined by 
\begin{align}
    w_{X,i}&={\rm max}\left(0, w_0+
{\rm log}\frac{E_i}{\sum_{j\in\rm v\,layers}E_{j}}\right)\\
w_{Y,i}&={\rm max}\left(0, w_0+
{\rm log}\frac{E_i}{\sum_{j\in\rm h\,layers}E_{j}}\right)
\end{align}
The ``h layers'' sums are over layers with horizontally aligned strips and the ``v layers'' sums are over layers with vertically aligned strips. The cutoff parameter $w_{0}$ is set to 4.0.

\begin{figure}[h!]
    \centering
    \includegraphics[width=0.495\textwidth]{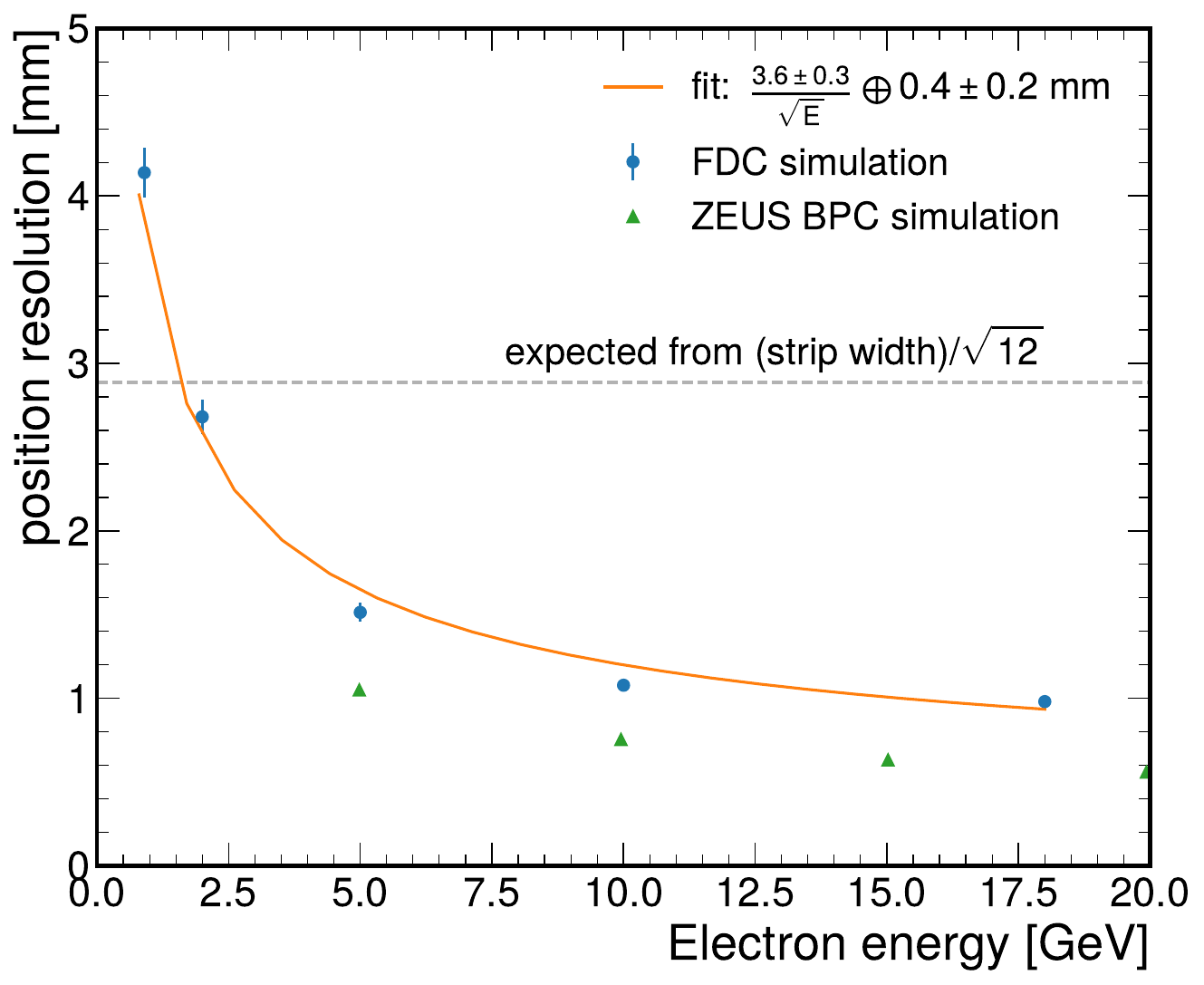}
    \caption{Position resolution of the FDC detector.  This is compared to the resolution determined via simulations for the ZEUS BPC~\cite{Surrow:1998su} (green triangles)}
    \label{fig:pos_res}
\end{figure}

Figure~\ref{fig:pos_res} shows the position resolution as a function of energy. For energies greater than 1 GeV, the position resolution is better than the strip width divided by $\sqrt{12}$. The resolution we obtained is poorer than the ZEUS BPC resolution. This difference can be partially explained by the smaller strip width (7.9 mm vs 10 mm), but it may also include components from algorithm tuning. 

\subsection{Kinematic Variable Reconstruction}
The resolution in Bjorken $x$ and $Q^2$ can be derived from those of the electron energy $\delta E'_e$ and of the electron polar angle $\delta \theta_e$ (using the small-angle approximation):
\begin{align}
    \frac{\delta Q^2}{Q^2}&\approx\frac{\delta E_e'}{E_e'} \oplus  \frac{2}{\pi-\theta_e}\delta \theta_e, \\
    \frac{\delta x}{x}&\approx\frac{1}{y}\frac{\delta E_e'}{E_e'}\oplus \left(\frac{x}{E_e/E_p}-1\right)\frac{2}{\pi-\theta_e} \delta\theta_e
    \label{eq:res_Q2}
\end{align}

In the non-divergent region (\textit{i.e.} $y>0.1$), the $Q^{2}$ resolution ranges from 4\% to 14\% depending on kinematics, whereas the $x$ resolution ranges from 10\% to 50\% with a strong $y$ dependence. To quantify these resolutions in the context of inclusive DIS measurements, we followed the EIC YR approach and calculated the corresponding purity and stability values. In this context, purity is determined by calculating the fraction of events reconstructed in a specific bin that were also generated in that bin (i.e., $P = N_{\mathrm{(rec,gen)}} / N_\mathrm{rec}$). Stability is calculated as the fraction of events generated in a specific bin that were also reconstructed in that bin (i.e., $S = N_{\mathrm{(rec,gen)}} / N_{\mathrm{gen}}$). Here, $N_{\mathrm{(rec,gen)}}$ represents the number of events where the electron is both generated and reconstructed in the same bin. For the generated events, we used the same \textsc{Pythia6} simulation as described in Section~\ref{sec:bkgrejection}.

\begin{figure}[h!]
    \centering
    \includegraphics[width=0.495\textwidth]{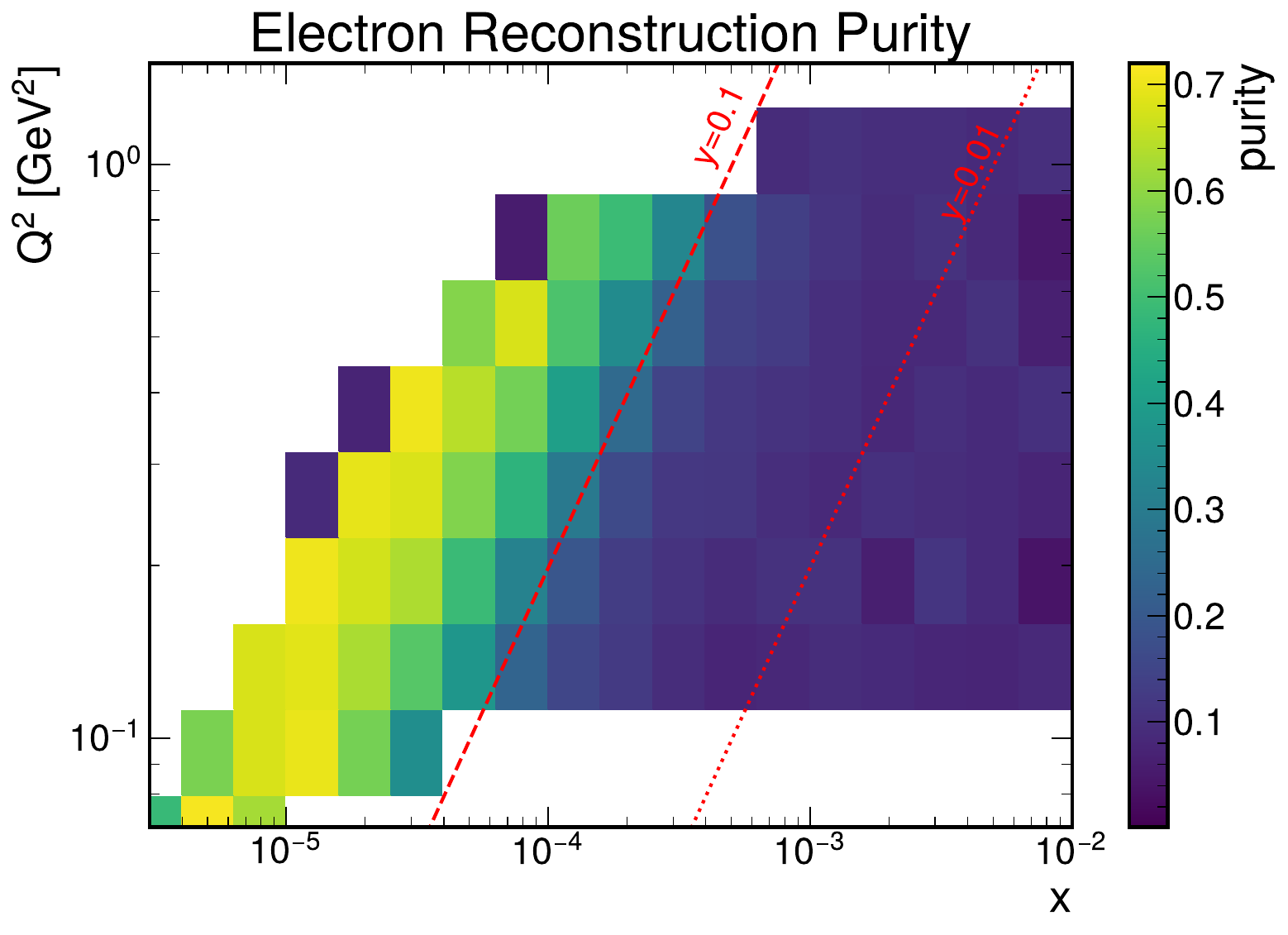}
    \includegraphics[width=0.495\textwidth]{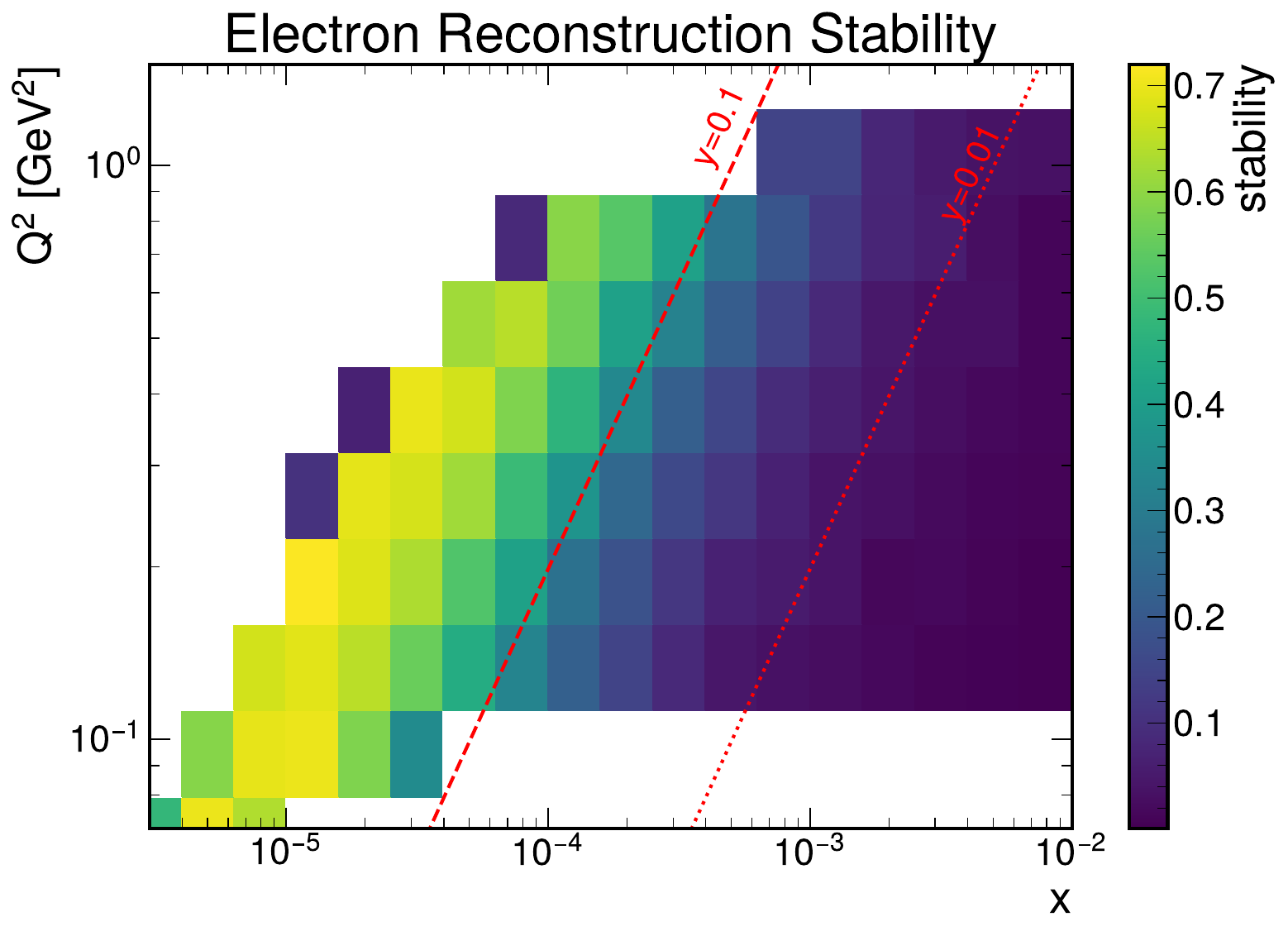}
    \caption{Purity (left) and Stability (right) for reconstruction of $x$ and $Q^{2}$ using the FDC and the electron method, for $ep$ collisions with 18$\times$ 275 GeV configuration.}
    \label{fig:purity}
\end{figure}

Figure~\ref{fig:purity} shows the resulting purity and stability plot for events with $-4.6<\eta<-3.6$. The plot has 5 bins per decade in both $x$ and $Q^{2}$, similar to the EIC Yellow Report~\cite{AbdulKhalek:2021gbh}. Purity and stability values above 50\% are observed for the phase-space covered with $y>0.1$, with some degradation at lower values, as expected when using the electron reconstruction method exclusively. It is anticipated that the performance for $0.01<y<0.1$ will improve by combining the electron and hadronic methods, potentially using machine-learning techniques~\cite{Arratia:2021tsq}.

A potential issue that can impact purity and stability is the electron beam's angular divergence, which can reach up to 200 $\mu$rad~\cite{eic_cdr}. Although this limits the kinematic reconstruction for electrons scattered at angles less than 10 mrad~\cite{eic_cdr}, the reconstruction of inclusive kinematic variables will not be compromised since the FDC acceptance begins at 20 mrad.

Overall, these studies demonstrate that the FDC design can provide sufficient resolution for measuring kinematic variables in the low $x$, low $Q^{2}$ region, thereby bridging the $Q^{2}$ gap.

\subsection{Shower-shape Examples}

\begin{figure*}[h!]
    \centering
  \includegraphics[width=.32\textwidth]{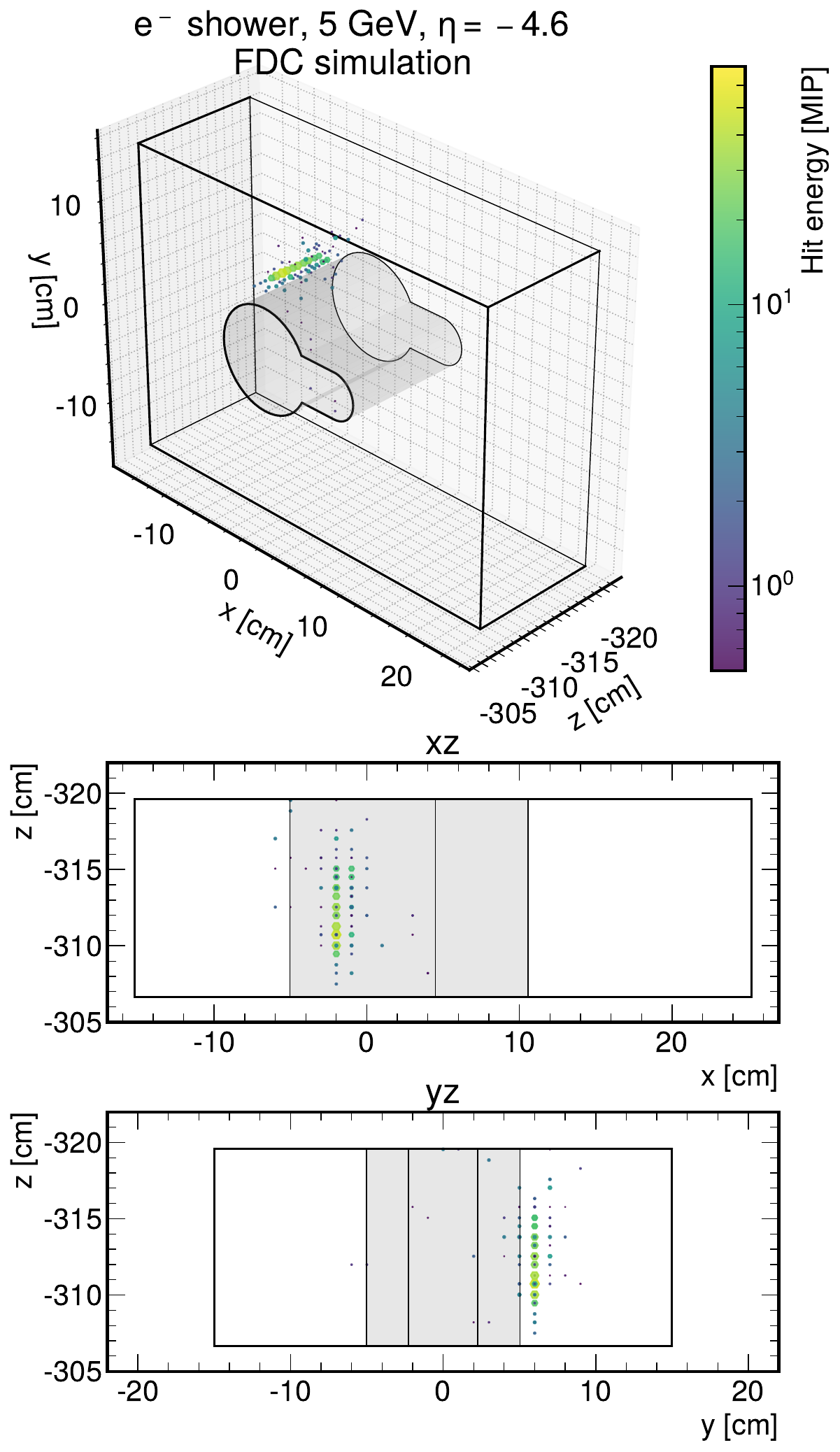}\includegraphics[width=.32\textwidth]{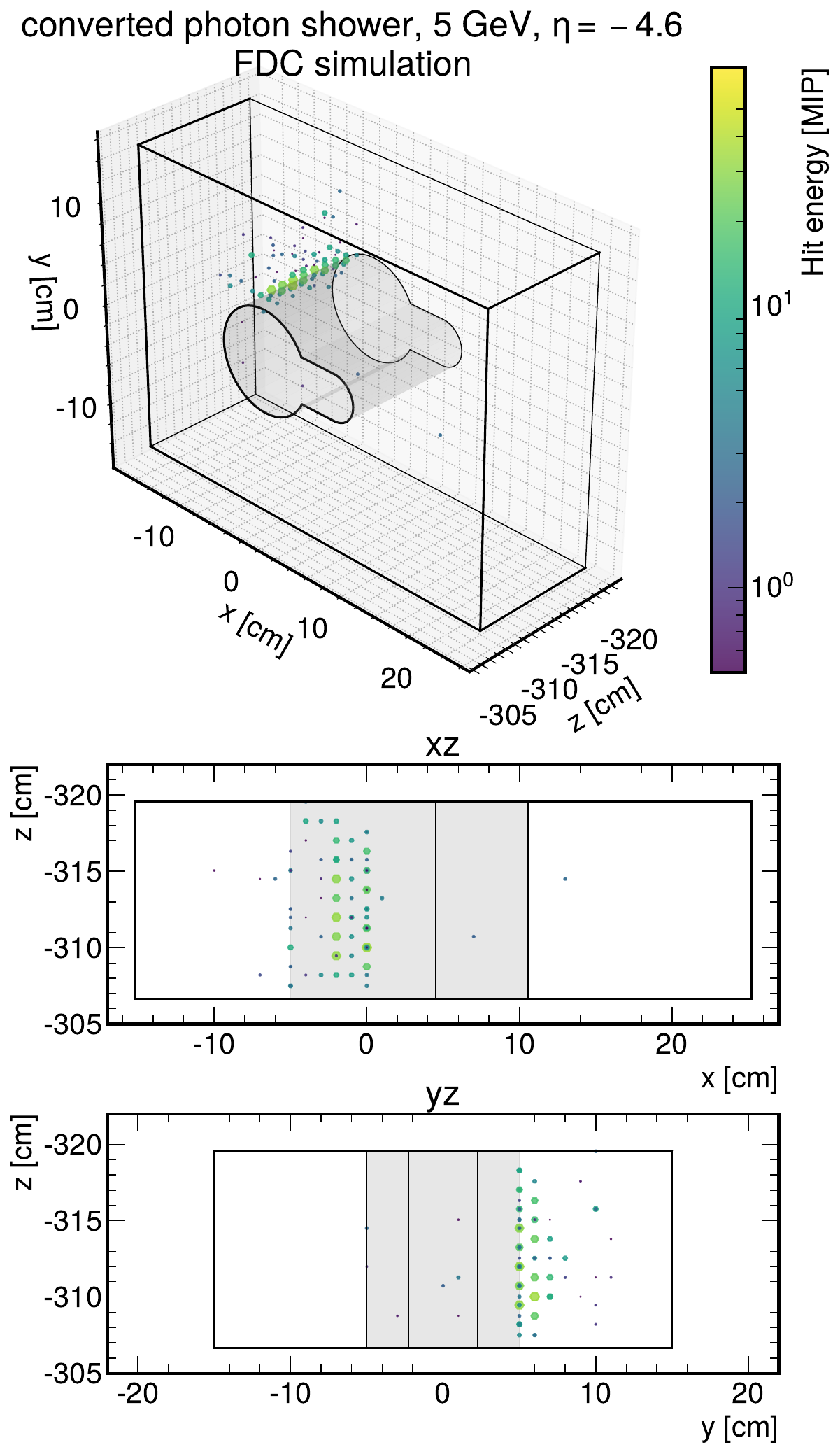}\includegraphics[width=.32\textwidth]{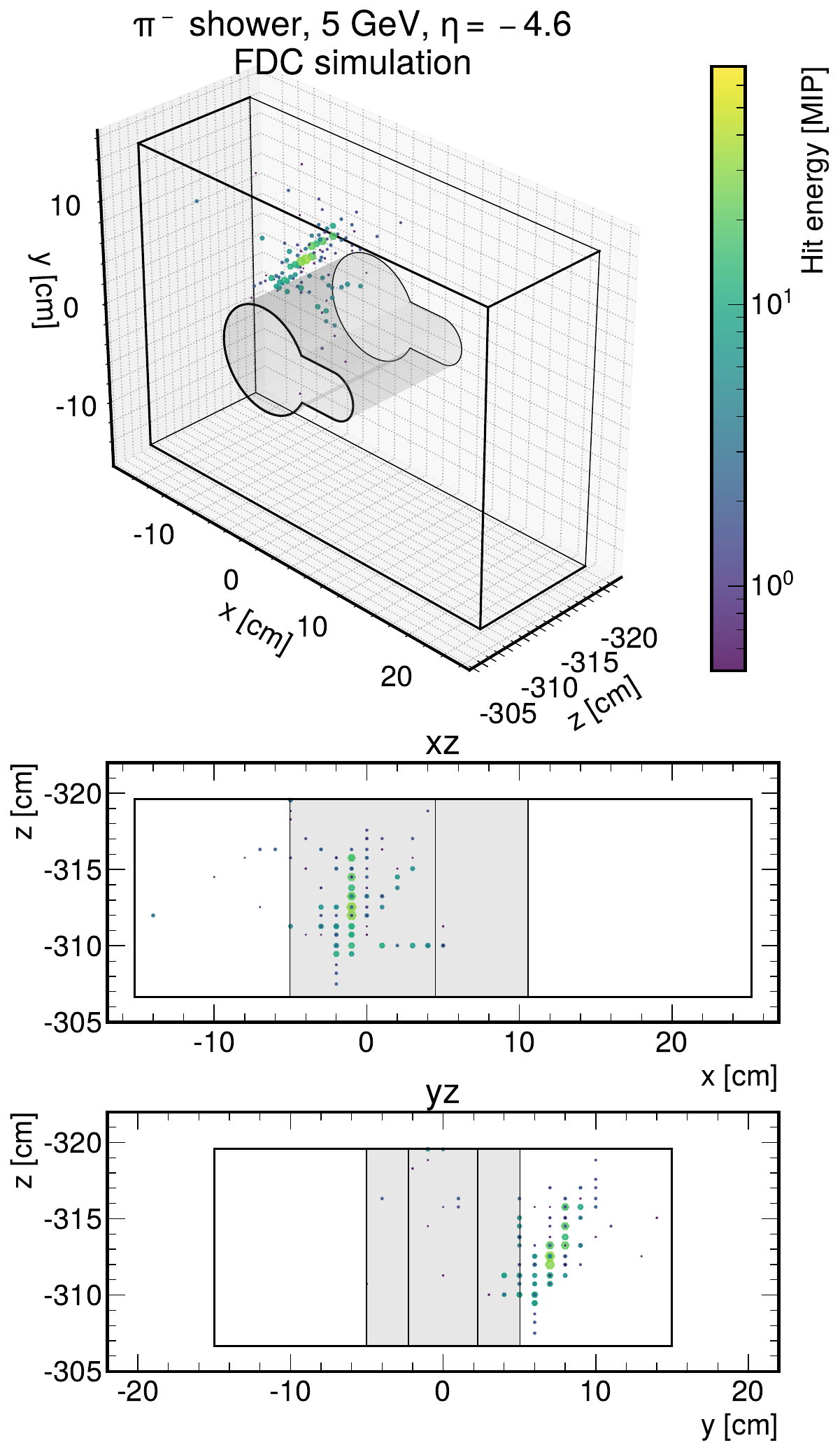}
    \caption{Examples of reconstructed showers are shown for electron (left), converted-photon (middle), and charged pion (right), with 5 GeV and $\eta=-4.6$. The displayed granularity is 10$\times$10 mm$^{2}$, which represents the expected effective granularity of the FDC after applying the strip-split algorithm~\cite{KOTERA2015158}. }
    \label{fig:showershape}
\end{figure*}

Figure~\ref{fig:showershape} shows 3D and projection views of example showers. The three scenarios depicted are: an electron reaching the FDC with no material in front of it (left), a photon that initiated showering in the beampipe (middle), and a $\pi^-$. Among the three cases, the electron without pre-showering produces the narrowest shower, while the pre-showering photon and the $\pi^-$ generate more irregular showers.

In terms of converted-electron tagging, the showers are expected to exhibit broader transverse profiles and anomalous longitudinal development. For hadron tagging, the fine segmentation will be primarily used to identify the starting point of the shower, which is more likely to be located at a deeper position compared to electrons. Moreover, the hadron showers will also have a different time development, with hits at later times compared to electron showers. 

This highlights the potential of 5D shower shape analysis for background rejection. In our future work, we intend to explore this potential by applying machine learning techniques to tag electrons/photons, hadron showers, and beam-gas background. Some promising emerging work, which can handle the complex geometry of the FDC, are point-cloud networks~\cite{ATL-PHYS-PUB-2022-040}.

\section{Summary and Conclusions}
\label{sec:conclusions}
We have outlined the design of a small electromagnetic calorimeter, the Few-Degree Calorimeter (FDC), which is designed to cover the range of $-4.6 < \eta < -3.6$. The primary objective of this detector is to tag electrons within the $Q^2$ range of 0.1 to 1.0 GeV$^2$, thus enabling future research on the transition to perturbative QCD and the gluon-saturation regime.

The FDC design we present here incorporates the latest advancements in SiPM-on-tile calorimetry to create a modern and improved version of the ZEUS Beam Pipe Calorimeter and H1 Very Low $Q^{2}$ calorimeter. The incorporation of high-granularity 5D shower measurements (position, time, and energy) offered by this technology holds great potential for background tagging.

In conclusion, this document presents the first design that has the potential to close the EIC $Q^2$ gap while maintaining a compact and cost-effective solution. Considering the larger crossing-angle envisioned for the second-interaction region at the EIC, which results in a larger hole in the crystal ECAL acceptance, this design may offer further opportunities for optimization for the EIC Detector 2.

\section*{Acknowledgements} \label{sec:acknowledgements}
We would like to extend our sincere appreciation to the members of the California EIC consortium for their valuable feedback on the design and physics that motivated the FDC, with special recognition to Oleg Tsai, Farid Salazar, and Zhongbo Kang. Additionally, we are grateful to Elke-Caroline Aschenauer for her guidance regarding the current acceptance of the crystal ECAL in ePIC. 

This work was supported by MRPI program of the University of California Office of the President, award number 00010100. This work was supposed by DOE grant award number DE-SC0022324. S.P also acknowledges support from the Jefferson Lab EIC Center Fellowship. M.A acknowledges support through DOE Contract No. DE-AC05-06OR23177 under which Jefferson Science Associates, LLC operates the Thomas Jefferson National Accelerator Facility.
\renewcommand\refname{Bibliography}
\bibliographystyle{utphys} 
\bibliography{bibio.bib} 

\providecommand{\href}[2]{#2}\begingroup\raggedright\begin{thebibliography}{10}

\bibitem{AbdulKhalek:2021gbh}
R.~Abdul~Khalek {\em et~al.}, ``{Science Requirements and Detector Concepts for
  the Electron-Ion Collider}: {EIC Yellow Report},''
  \href{http://dx.doi.org/10.1016/j.nuclphysa.2022.122447}{{\em Nucl. Phys. A}
  {\bfseries 1026} (2022) 122447},
  \href{http://arxiv.org/abs/2103.05419}{{\ttfamily arXiv:2103.05419
  [physics.ins-det]}}.

\bibitem{managerie}
``{EIC Detector Geometry Interaction Point 6 (version January 2023)
  ePIC-Detector.skp}.''
  \url{https://eic.jlab.org/Geometry/Detector/Detector-20230108185912.html }.
\newblock Accessed: 2023-06-18.

\bibitem{ELKEprivate}
E.-C. Aschenauer, ``Private communication.'' Unpublished.

\bibitem{GDI}
``{Global Detector Integration Working Group Meeting, Low $Q^2$ Coverage of
  ePIC (Monday 6 Feb 2023)}.'' \url{https://indico.bnl.gov/event/18161/}.
\newblock Accessed: 2023-06-15.

\bibitem{Surrow:1998su}
B.~Surrow, ``{Measurement of the proton structure function $F_2$ at low $Q^2$
  and very low x with the ZEUS beam pipe calorimeter at HERA},''
  \href{http://dx.doi.org/10.1007/s1010599c0002}{{\em Eur. Phys. J. direct}
  {\bfseries 1} no.~1, (1999) 2}.

\bibitem{Stellberger_2003}
A.~Stellberger, J.~Ferencei, F.~Kriv{\'{a}}{\v{n}}, K.~Meier, O.~Niedermaier,
  O.~Nix, K.~Schmitt, J.~{\v{S}}palek, J.~Stiewe, and M.~Weber, ``The {VLQ}
  calorimeter of {H1} at {HERA}: a highly compact device for measurements of
  electrons and photons under very small scattering angles,''
  \href{http://dx.doi.org/10.1016/j.nima.2003.08.111}{{\em Nuclear Instruments
  and Methods in Physics Research Section A: Accelerators, Spectrometers,
  Detectors and Associated Equipment} {\bfseries 515} no.~3, (Dec, 2003)
  543--562}. \url{https://doi.org/10.1016%2Fj.nima.2003.08.111}.

\bibitem{PhysRevLett.86.596}
A.~M. Sta\ifmmode~\acute{s}\else \'{s}\fi{}to, K.~Golec-Biernat, and
  J.~Kwieci\ifmmode~\acute{n}\else \'{n}\fi{}ski, ``Geometric scaling for the
  total ${\mathit{\ensuremath{\gamma}}}^{*}\mathit{p}$ cross section in the low
  $\mathit{x}$ region,''
  \href{http://dx.doi.org/10.1103/PhysRevLett.86.596}{{\em Phys. Rev. Lett.}
  {\bfseries 86} (Jan, 2001) 596--599}.
  \url{https://link.aps.org/doi/10.1103/PhysRevLett.86.596}.

\bibitem{H1:2005dtp}
{\bfseries H1} Collaboration, A.~Aktas {\em et~al.}, ``{Elastic J/psi
  production at HERA},''
  \href{http://dx.doi.org/10.1140/epjc/s2006-02519-5}{{\em Eur. Phys. J. C}
  {\bfseries 46} (2006) 585--603},
  \href{http://arxiv.org/abs/hep-ex/0510016}{{\ttfamily arXiv:hep-ex/0510016}}.

\bibitem{ZEUS:2007iet}
{\bfseries ZEUS} Collaboration, S.~Chekanov {\em et~al.}, ``{Exclusive $\rho^0$
  production in deep inelastic scattering at HERA},''
  \href{http://dx.doi.org/10.1186/1754-0410-1-6}{{\em PMC Phys. A} {\bfseries
  1} (2007) 6}, \href{http://arxiv.org/abs/0708.1478}{{\ttfamily
  arXiv:0708.1478 [hep-ex]}}.

\bibitem{PhysRevLett.100.022303}
H.~Kowalski, T.~Lappi, and R.~Venugopalan, ``Nuclear enhancement of universal
  dynamics of high parton densities,''
  \href{http://dx.doi.org/10.1103/PhysRevLett.100.022303}{{\em Phys. Rev.
  Lett.} {\bfseries 100} (Jan, 2008) 022303}.
  \url{https://link.aps.org/doi/10.1103/PhysRevLett.100.022303}.

\bibitem{MechanicalDesign}
``{EEEMCal, (Electron Ion Collider - EIC) Mechanical design and Integration.
  Version 1.2 }.''
  \url{https://wiki.jlab.org/cuawiki/images/e/e5/EEEMCAL_Mechanical_design_2.pdf
  }.
\newblock Accessed: 2023-06-18.

\bibitem{ZEUS:1997etp}
{\bfseries ZEUS} Collaboration, J.~Breitweg {\em et~al.}, ``{Measurement of the
  proton structure function $F_2$ and $\sigma_{tot}^{\gamma* p}$ at low Q$^2$
  and very low x at HERA},''
  \href{http://dx.doi.org/10.1016/S0370-2693(97)00905-2}{{\em Phys. Lett. B}
  {\bfseries 407} (1997) 432--448},
  \href{http://arxiv.org/abs/hep-ex/9707025}{{\ttfamily arXiv:hep-ex/9707025}}.

\bibitem{Sefkow:2015hna}
F.~Sefkow, A.~White, K.~Kawagoe, R.~P\"oschl, and J.~Repond, ``{Experimental
  Tests of Particle Flow Calorimetry},''
  \href{http://dx.doi.org/10.1103/RevModPhys.88.015003}{{\em Rev. Mod. Phys.}
  {\bfseries 88} (2016) 015003},
  \href{http://arxiv.org/abs/1507.05893}{{\ttfamily arXiv:1507.05893
  [physics.ins-det]}}.

\bibitem{CALICE:2017sis}
{\bfseries CALICE} Collaboration, J.~Repond {\em et~al.}, ``{Construction and
  Response of a Highly Granular Scintillator-based Electromagnetic
  Calorimeter},'' \href{http://dx.doi.org/10.1016/j.nima.2018.01.016}{{\em
  Nucl. Instrum. Meth. A} {\bfseries 887} (2018) 150--168},
  \href{http://arxiv.org/abs/1707.07126}{{\ttfamily arXiv:1707.07126
  [physics.ins-det]}}.

\bibitem{KOTERA2015158}
K.~Kotera, D.~Jeans, A.~Miyamoto, and T.~Takeshita, ``A novel strip energy
  splitting algorithm for the fine granular readout of a scintillator strip
  electromagnetic calorimeter,''
  \href{http://dx.doi.org/https://doi.org/10.1016/j.nima.2015.04.001}{{\em
  Nuclear Instruments and Methods in Physics Research Section A: Accelerators,
  Spectrometers, Detectors and Associated Equipment} {\bfseries 789} (2015)
  158--164}.
  \url{https://www.sciencedirect.com/science/article/pii/S0168900215004544}.

\bibitem{Dong:2018hvs}
{\bfseries CEPC calorimeter working group} Collaboration, M.~Y. Dong, ``{R\&D
  of the CEPC scintillator-tungsten ECAL},''
  \href{http://dx.doi.org/10.1088/1748-0221/13/03/C03024}{{\em JINST}
  {\bfseries 13} no.~03, (2018) C03024}.

\bibitem{Niu:2020iln}
Y.~Niu {\em et~al.}, ``{Design of Sc-ECAL prototype for CEPC and performance of
  first two layers},''
  \href{http://dx.doi.org/10.1088/1748-0221/15/05/C05036}{{\em JINST}
  {\bfseries 15} no.~05, (2020) C05036},
  \href{http://arxiv.org/abs/2002.01809}{{\ttfamily arXiv:2002.01809
  [physics.ins-det]}}.

\bibitem{Arratia:2022quz}
M.~Arratia {\em et~al.}, ``{A high-granularity calorimeter insert based on
  SiPM-on-tile technology at the future Electron-Ion Collider},''
  \href{http://dx.doi.org/10.1016/j.nima.2022.167866}{{\em Nucl. Instrum. Meth.
  A} {\bfseries 1047} (2023) 167866},
  \href{http://arxiv.org/abs/2208.05472}{{\ttfamily arXiv:2208.05472
  [physics.ins-det]}}.

\bibitem{Arratia:2023rdo}
M.~Arratia, L.~Garabito~Ruiz, J.~Huang, S.~J. Paul, S.~Preins, and
  M.~Rodriguez, ``{Studies of time resolution, light yield, and crosstalk using
  SiPM-on-tile calorimetry for the future Electron-Ion Collider},''
  \href{http://dx.doi.org/10.1088/1748-0221/18/05/P05045}{{\em JINST}
  {\bfseries 18} no.~05, (2023) P05045},
  \href{http://arxiv.org/abs/2302.03646}{{\ttfamily arXiv:2302.03646
  [physics.ins-det]}}.

\bibitem{Doses}
``Radiation dose and neutron flux from minimum-bias e+p events at 18x275 gev
  )..''
  \url{https://wiki.bnl.gov/EPIC/index.php?title=Radiation_Doses#Radiation_Dose_and_Neutron_Flux_from_Minimum-Bias_e+p_events_@_18x275_GeV}.
\newblock Accessed: 2023-06-18.

\bibitem{Frank:2014zya}
M.~Frank, F.~Gaede, C.~Grefe, and P.~Mato, ``{DD4hep: A Detector Description
  Toolkit for High Energy Physics Experiments},''
  \href{http://dx.doi.org/10.1088/1742-6596/513/2/022010}{{\em J. Phys. Conf.
  Ser.} {\bfseries 513} (2014) 022010}.

\bibitem{GEANT4:2002zbu}
{\bfseries GEANT4} Collaboration, S.~Agostinelli {\em et~al.}, ``{GEANT4--a
  simulation toolkit},''
  \href{http://dx.doi.org/10.1016/S0168-9002(03)01368-8}{{\em Nucl. Instrum.
  Meth. A} {\bfseries 506} (2003) 250--303}.

\bibitem{Monteiro:1998bi}
T.~Monteiro, \href{http://dx.doi.org/10.3204/PUBDB-2016-02635}{{\em {Study of
  exclusive electroproduction of $\rho^0$ mesons at low $Q^2$ using the ZEUS
  Beam Pipe Calorimeter at HERA}}}.
\newblock PhD thesis, Hamburg U., 1998.

\bibitem{Arratia:2021tsq}
M.~Arratia, D.~Britzger, O.~Long, and B.~Nachman, ``{Reconstructing the
  kinematics of deep inelastic scattering with deep learning},''
  \href{http://dx.doi.org/10.1016/j.nima.2021.166164}{{\em Nucl. Instrum. Meth.
  A} {\bfseries 1025} (2022) 166164},
  \href{http://arxiv.org/abs/2110.05505}{{\ttfamily arXiv:2110.05505
  [hep-ex]}}.

\bibitem{eic_cdr}
F.~Willeke and J.~Beebe-Wang, ``{Electron Ion Collider Conceptual Design Report
  2021},'' \href{http://dx.doi.org/10.2172/1765663}{ (2, 2021) }.
  \url{https://www.osti.gov/biblio/1765663}.

\bibitem{ATL-PHYS-PUB-2022-040}
{\bfseries ATLAS} Collaboration, ``{Point Cloud Deep Learning Methods for Pion
  Reconstruction in the ATLAS Experiment},'' tech. rep., CERN, Geneva, 2022.
\newblock \url{https://cds.cern.ch/record/2825379}.
\newblock All figures including auxiliary figures are available at
  https://atlas.web.cern.ch/Atlas/GROUPS/PHYSICS/PUBNOTES/ATL-PHYS-PUB-2022-040.

\end{thebibliography}\endgroup

\appendix
\end{document}